\documentclass[prl,aps,twocolumn,superscriptaddress,amsmath,longbibliography]{revtex4-2}
\usepackage{graphicx,color,hyperref,xcolor}
\usepackage{soul}
\usepackage{amsfonts}
\usepackage{amssymb}
\usepackage{amsmath}
\usepackage{amsthm}
\usepackage{bm}
\usepackage{braket}
\usepackage{enumerate}

\usepackage{CJKutf8}

\begin{document}
\begin{CJK}{UTF8}{gbsn}

\title{Driven-dissipative superconductivity in moiré heterostructure without attraction}

\author{Tsung-Sheng Huang}
\affiliation{Joint Quantum Institute, University of Maryland, College Park, MD 20742, USA}
\affiliation{ICFO-Institut de Ciencies Fotoniques, The Barcelona Institute of Science and Technology, 08860 Castelldefels (Barcelona), Spain}

\author{Ataç Imamoglu}
\affiliation{Institute of Quantum Electronics, ETH Zürich, 8093 Zürich, Switzerland}

\author{Mohammad Hafezi}
\affiliation{Joint Quantum Institute, University of Maryland, College Park, MD 20742, USA}

\author{Sebastian Diehl}
\affiliation{Institute for Theoretical Physics, University of Cologne, 50937 Cologne, Germany}

\date{\today}

\begin{abstract}

Dissipative preparation of quantum order offers a route to superconductivity that does not rely on enhancing attractive interactions. Here we propose a driven–dissipative protocol to prepare superconductivity as a stationary state of a two-dimensional moiré heterostructure. The key ingredient is a bilayer moiré platform in which the layer degree of freedom acts as a pseudospin, allowing the pseudospin structure required for pairing to be implemented through optically induced spatial operations. 
This preparation scheme requires local dissipation, which we show to arises naturally from weakly dispersive bosonic modes in the heterostructure.
In contrast, in the opposite regime of collective dissipation, the same platform exhibits an early-time superradiant burst. Our results establish driven–dissipative moiré heterostructures as a promising platform for preparing superconductivity, while also revealing a connection between steady-state pairing and transient superradiance.

\end{abstract}

\maketitle
\end{CJK}

Beyond the search for new materials~\cite{dagotto1994correlated,orenstein2000advances,lee2006doping}, efforts to realize superconducting (SC) order have also motivated light-based nonequilibrium approaches.
One established strategy is to use optical driving to strengthen the attractive interaction between electrons~\cite{sentef2016theory,kennes2017transient,eckhardt2024theory}.
While transient SC has been experimentally demonstrated in this context~\cite{fausti2011light,mitrano2016possible,budden2021evidence}, the extension of this approach toward long-lived order remains an open challenge.
Instead of manipulating the attraction, a promising alternative route is to use external drives to engineer dissipation, so that the stability of SC is built into the mechanism itself~\cite{diehl2010dissipation}.
These dissipators are designed to remove excitations above the target order, guiding the system toward the corresponding quantum state in steady state~\cite{Diehl2008,Verstraete2009,Weimer_2010,Barreiro_2011,diehl2011np,Krauter2011,harrington2022engineered,mi2024stable}.

A central challenge in implementing this strategy is the need to engineer drive and dissipation with spatial and spin structures dictated by the target pair wavefunction. 
Specifically, achieving simultaneous control over both is typically experimentally demanding. 
One natural way to ease this difficulty is to consider pseudospins that arise from positional degrees of freedom, thereby reducing the problem essentially to spatial control alone. 
In this context, bilayer materials are particularly promising, as the layer degree of freedom naturally serves as such a pseudospin. 
In particular, recent developments have demonstrated the ability to optically switch the layer pseudospin of particles in these bilayers~\cite{leisgang2020giant,barre2022optical}, providing access to the spin-flip dissipative operations required for certain pairing symmetries.

\begin{figure}[t]
\centering
\includegraphics[width=0.9\linewidth]{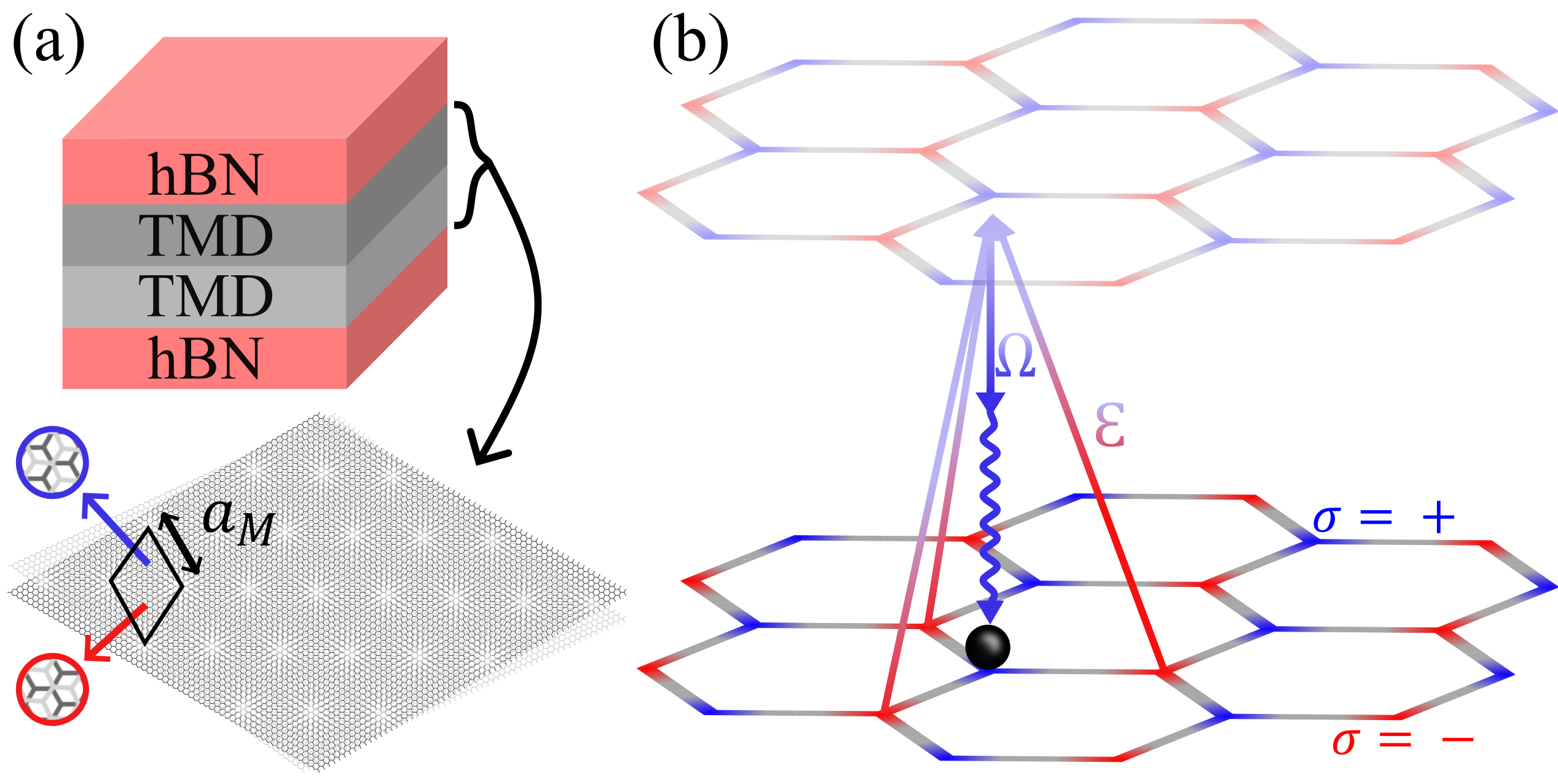}
\caption{
\textbf{Illustration of the driven-dissipative system}.
(a)
Top: The moiré heterostructure formed by a TMD homobilayer sandwiched by hBNs.
Bottom: Moiré pattern from TMDs, giving two high symmetry points (red and blue) related to each other by a top-down inversion, which offers pseudospin degrees of freedom ($ \sigma = \pm$).
(b) Engineered dissipation for the preparation of superconductivity.
A Raman transition constituted by two external drives $\mathcal{E}$ and $\Omega$ (straight arrows) is implemented to access the emission of bosonic modes in hBN (curly arrow).
These processes, with the auxiliary lattice (top hexagons) off-resonantly involved, together lead to movement of a fermion (black dot) in the target lattice (bottom hexagons), with both lattices corresponding to the high symmetry points of the moiré TMD.
Pairs of fermions are induced by the dissipative processes associated to such an incoherent motion.
}
\label{Fig_1}
\end{figure}

In this work, we propose to dissipatively prepare SC in a two-dimensional heterostructure of stacked atomically thin layers [Fig.~\ref{Fig_1}(a)].
The central part of this device consists of a moiré transition metal dichalcogenide (TMD) homobilayer, in which the charges that eventually form the SC pairs reside.
These charges live on a bipartite superlattice whose two sublattices are microscopically associated with different layers, thereby defining a layer pseudospin.
Consequently, a flip of this pseudospin is equivalent to nearest-neighbor charge transfer, which enters the dissipative processes that ultimately drive pairing [Fig.~\ref{Fig_1}(b)].

In our design, such dissipators are engineered through the combination of Raman transitions and bosonic-mode emission. 
The Raman processes induce couplings between moiré unit cells and layers, enabling access to the required structure dictated by the target pair wavefunction. 
In addition to these drives, incoherent processes are required to bring the system to a steady state, here provided by the emission into bosonic modes.

\begin{figure}
\centering
\includegraphics[width=\linewidth]{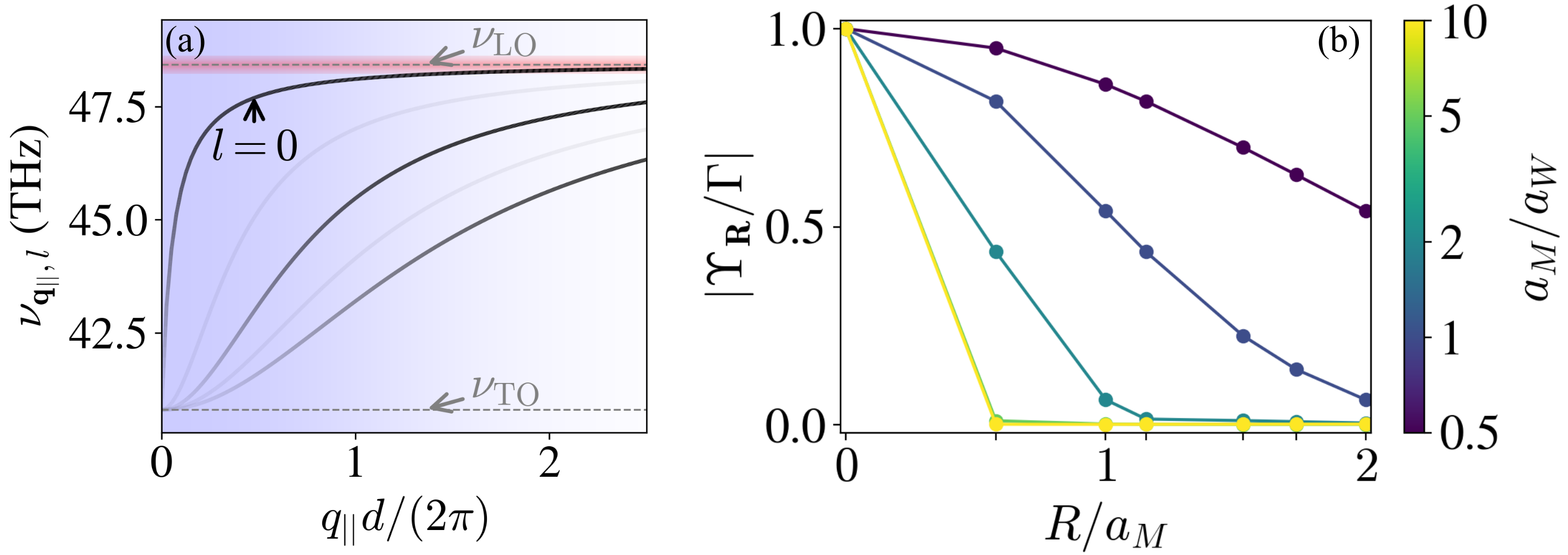}
\caption{
\textbf{Properties of HPPs in hBN}.
(a) Dispersion in HPPs.
Opacity indicates the associated in-plane electric field strength $|E_n^{||}|$, which is involved in the coupling to fermions.
$l=0$ labels the topmost mode arising from out-of-plane confinement (see Methods).
The red shaded region marks the target linewidth centered around the longitudinal optical phonon frequency $\nu_{\mathrm{LO}}$.
The blue shading indicates the relative weight of momenta contributing to dissipation, set by the width of the fermionic Wannier orbital.
(b) Dissipation matrix $\Upsilon_{\bm{R}}$ mediated by HPPs at various distances $R$ and moiré periods $a_M$.
In this panel, we set the fermionic Wannier orbital $a_W = 2$ nm~\cite{karni2022structure}, Raman detuning $\nu = 48.42$ THz~\cite{feng2026color}, and hBN thickness $d = 5$ nm~\cite{feng2026color}. 
}
\label{Fig_2}
\end{figure}

We further find that interference among the emitted modes disfavors the emergence of SC as a steady state.
This observation precludes the use of photons as the loss mediator, because of their nonlocal character on the superlattice scale.
As a solution, we sandwich the TMD layers in the heterostructure with hexagonal boron nitride (hBN) sheets and exploit the hyperbolic phonon-polaritons (HPPs) therein.
These modes possess a nearly nondispersive sector [Fig.~\ref{Fig_2}(a)], and the resulting locality suppresses their mutual interference.
Combining HPPs with the Raman processes, we numerically consolidate that the dissipative dynamics brings the system to a BCS paired state at long times.

Finally, our design offers the tunability to switch the relevant emitted modes between photons and HPPs, and hence to control the degree of interference in the dissipators.
In the complementary regime where photons mediate the emission, we find the emergence of collective radiation, manifested as an early-time burst in the photoemission rate.
Our results therefore show that driven-dissipative moiré heterostructures can host both steady-state superconductivity and transient superradiance, unifying these seemingly distinct many-body phenomena within a common framework in which interference in dissipation acts as a tunable control parameter.

\textit{\textbf{SC without attractive interaction.}} --- 
We  engineer a Lindblad dynamics on a 2D bipartite lattice, where the time evolution of the system density operator $\hat{\rho}$ is governed by
\begingroup
\thinmuskip=1.5mu
\medmuskip=2mu
\thickmuskip=2.5mu
\begin{equation}
\label{eq:Lindblad_model_SC}
\begin{aligned}
\partial_t\hat{\rho}
+i
[
\hat{\mathcal{H}}
,
\hat{\rho}
]
=
\Gamma
\sum_{\bm{r},\sigma}
\left[
\hat{T}_{\bm{r},\sigma}
\hat{\rho}
\hat{T}_{\bm{r},\sigma}^\dagger
-
\left\{
\frac{\hat{\rho}}{2}
,
\hat{T}_{\bm{r},\sigma}^\dagger
\hat{T}_{\bm{r},\sigma}
\right\}
\right]
.
\end{aligned}
\end{equation}
\endgroup
Here, the bipartite (honeycomb) lattice is parametrized by the unit cell index $\bm{r}$ and the sublattice-layer pseudospin $\sigma =\pm$.
$\hat{\mathcal{H}}$ denotes a Hamiltonian, and the local Lindblad jump operators $\hat{T}_{\bm{r},\sigma}$ describe the effective system-bath coupling to be engineered, with the corresponding dissipation rate given by $\Gamma$. 
Our approach exploits \textit{dark states} $|D\rangle$, which are common zero modes of all Lindblad operators, and eigenstates of the Hamiltonian: 
\begin{equation}
\label{eq:dark_state_condition}
\hat{T}_{\bm{r},\sigma}|D\rangle=0
,\;\forall \,(\bm{r},\sigma)
;\quad
\hat{\mathcal{H}} = E|D\rangle\, .
\end{equation}
Therefore, dark states are stationary states of the Lindblad evolution. 
Here we are interested in situations where the dark state is the \textit{unique} solution of the Lindblad equation: 
In this case, the evolution steers the system density matrix into the dark state, independently of the initial state~\cite{Diehl2008,Verstraete2009}. 
A pure state of matter is then reached by a driven-dissipative evolution. 

Our specific goal is to design $\hat{T}_{\bm{r},\sigma}$ such that $|D\rangle$ corresponds to the BCS superconducting state $|\text{BCS}\rangle$~\cite{diehl2010dissipation,diehl2011np}. 
The mechanism involves two steps [Fig.~\ref{Fig_1}(b)]: fermions are driven into an auxiliary superlattice system and decay back into the target lattice via the phonon bath.
The Lindblad operators describing such net process take the form:
\begingroup
\thinmuskip=1mu
\medmuskip=1.5mu
\thickmuskip=2mu
\begin{equation}
\label{eq:jump_general}
\hat{T}_{\bm{r},\sigma} 
\equiv 
\hat{C}_{\bm{r},\sigma}^\dagger
\hat{A}_{\bm{r},\sigma}
,\;\;\;
\begin{bmatrix}
\check{C}_{\bm{r}}^\dagger
\\
\check{A}_{\bm{r}}
\end{bmatrix}
\equiv
\sum_{\bm{r}'}
\begin{bmatrix}
\check{v}({\bm{r}-\bm{r}'}) \cdot \check{f}_{\bm{r}'}^\dagger
\\
\check{u}({\bm{r}-\bm{r}'}) \cdot \check{f}_{\bm{r}'}
\end{bmatrix}
,
\end{equation}
\endgroup
where $\check{C}_{\bm{r}}^\dagger\equiv(\hat{C}_{\bm{r},+}^\dagger\,,\, \hat{C}_{\bm{r},-}^\dagger)^\text{T}$ and $\check{A}_{\bm{r}}\equiv(\hat{A}_{\bm{r},+}\,,\, \hat{A}_{\bm{r},-})^\text{T}$ describe the transition of fermions between target and auxiliary systems via optical drives and interaction with the bath, respectively.
Here we focus on the parameter regime where both the auxiliary lattice and the bath can be adiabatically eliminated, such that these operators involve solely the fermions within the target lattice, with annihilator $\check{f}_{\bm{r}'}\equiv(\hat{f}_{\bm{r},+}\,,\,\hat{f}_{\bm{r},-})^\text{T}$, which is considered spinless (yet equipped with pseudospin $\sigma\in\pm$) as appropriate for our implementation below. 
The matrices $\check{v}({\bm{r}-\bm{r}'})$ and $\check{u}({\bm{r}-\bm{r}'})$ encode the strength as well as spatial and pseudospin symmetries of the driving and decay process, which we aim to be translationally invariant;  their controlled engineering is the main target of our implementation proposal.

Before presenting the explicit design, we offer some intuition for why a BCS state naturally emerge as a dark state. 
By definition, such a state can be understood as the vacuum for Bogoliubov excitations $\hat{\gamma}_{\bm{k},\sigma} |\text{BCS}\rangle = 0$, where $\bm{k}$ is the conjugate momentum of $\bm{r}$, and $|\text{BCS}\rangle$ denotes the coherent state of pairs.
$\hat{\gamma}_{\bm{k},\sigma}$ obey the fermionic algebra, which, given proper choice of amplitudes in Eq.~\eqref{eq:jump_general}, can be satisfied with the following form (in its spinor representation $\check{\gamma}_{\bm{k}}$):
\begin{equation}
\label{eq:Bogoliubov_op}
\check{\gamma}_{\bm{k}} 
= 
\check{A}_{\bm{k}} 
+ 
\check{C}_{-\bm{k}}^\dagger 
=
\check{u}(\bm{k})
\cdot
\check{f}_{\bm{k}}
+ 
\check{v}(\bm{k})
\cdot
\check{f}^\dagger_{-\bm{k}}
,
\end{equation}
where $\check{A}_{\bm{k}}$ is the Fourier transform of $\check{A}_{\bm{r}}$ and similarly for $\check{C}_{\bm{k}}^\dagger$ (see Methods).
The Bogoliubov vacuum condition within such a choice directly yields $\check{A}_{\bm{r}}|\text{BCS}\rangle = - \check{C}_{\bm{r}}^\dagger|\text{BCS}\rangle$ in the real space domain, which, together with Pauli exclusion $(\hat{C}_{\bm{r},\sigma}^\dagger)^2 = 0$, leads to Eq.~\eqref{eq:dark_state_condition}, thereby justifying $|\text{BCS}\rangle$ as a dark state.
More precisely, as the jump operators Eq.~\eqref{eq:jump_general} preserve fermion number, the unique dark state is the BCS state projected to a definite pair-number $N$: $|D\rangle = |\text{BCS},N\rangle  \sim (\hat{B}^\dagger)^{N} |0\rangle$ ($|0\rangle$ denotes the fermion vacuum), which approaches the coherent state $|\text{BCS}\rangle$ in the thermodynamic limit.
Here the pair creation operator $\hat{B}^\dagger$ possesses the following explicit form:
\begin{eqnarray}
\label{eq:hatB}
\hat B^\dagger 
&=&  
\sum_{\bm{r},\bm{r'}}    
\check{f}_{\bm{r}}^\dagger 
\cdot
\check{\phi}(\bm{r}- \bm{r}')
\cdot
\check{f}_{\bm{r}'}^\dagger
,\quad 
\\
\check{\phi} (\bm{r} - \bm{r}') 
&=& 
\sum_{\bm{r}''}
\check{u}^{-1}(\bm{r} - \bm{r}'')
\check{v}(\bm{r}'' - \bm{r}') 
= 
- \check{\phi}^\text{T}(\bm{r}' - \bm{r})
\nonumber 
\end{eqnarray}
in terms of $\check{u}$ and $\check{v}$; they have to be constructed such that the pair wavefunction is totally antisymmetric. 
Note we can set $\hat{\mathcal H} = 0$ in this discussion -- superconducting order arises based on a purely driven-dissipative mechanism, in the absence of any attractive force. 
In our implementation below, a Hamiltonian is induced as a Lamb shift, which possesses $|\text{BCS},N\rangle$ as its eigenstate and therefore does not modify the dark state solution.

\textit{\textbf{BCS wave function design.}} --- 
We now specify the concrete form of the amplitudes in Eq.~\eqref{eq:jump_general} which we design in our implementation proposal below:
\begin{eqnarray}
\label{eq:design}
u_{\sigma,\sigma'}(\bm{r} - \bm{r}')
&=&
\delta_{\sigma,\sigma'}\delta_{\bm{r},\bm{r}'},
\\
v_{\sigma,\sigma'}(\bm{r}- \bm{r}') 
&=&
(i \check{\sigma}^y)_{\sigma,\sigma'} 
\delta_{\langle \bm{R}_{\bm{r}',-\sigma},\bm{R}_{\bm{r},\sigma}\rangle}
\nonumber
,
\end{eqnarray}
where $\check{\sigma}$ denotes the Pauli matrix, and $\bm{R}_{\bm{r},\sigma}$ is the position of site $(\bm{r},\sigma)$.
$\delta_{\langle \bm{R}_{\bm{r},\sigma},\bm{R}_{\bm{r}',\sigma'}\rangle}$ equals 1 for nearest neighbors and zero otherwise; for hexagonal lattice considered below, such a factor also dictates $\sigma \neq \sigma'$ [Fig.~\ref{Fig_1}(b)].
Here $\check{u}(\bm{r})$ acts locally in real space and preserves the pseudospin projection, hence $\hat A_{\bm{r},\sigma} = \hat f_{\bm{r},\sigma}$. 
Conversely, $\check{v}(\bm{r})$ is exclusively inter-sublattice, thereby flipping the pseudospin. 
Notably, although the $(i \check{\sigma}^y)$ structure in $\check{v}(\bm{r})$ suggests its implementation require an antisymmetric drive, an alternative with symmetric one is actually possible (see Methods).

Eq.~\eqref{eq:design} yields a jump operator that removes a fermion from a site $\bm{R}_{\bm{r},\sigma}$ and creates a symmetric superposition at nearest neighbors, which are opposite in pseudospin.
\begin{equation}
\label{eq:jump_design}
\hat{T}_{\bm{r},\sigma}
=
\sum_{\bm{r}'}
\hat{f}_{\bm{r}',-\sigma}^\dagger 
\delta_{\langle \bm{R}_{\bm{r}',-\sigma},\bm{R}_{\bm{r},\sigma}\rangle}
\hat{f}_{\bm{r},\sigma}
.
\end{equation}
The resulting BCS state is characterized by the pair wavefunction $\phi_{\sigma\sigma'}(\bm{r}- \bm{r}')  = v_{\sigma\sigma'}(\bm{r}- \bm{r}')$.
It is quasi-local, corresponding to well-localized Cooper pairs with support exclusively over nearest neighbours, akin to strong coupling superconductivity. 
The dissipatively induced wavefunction is properly antisymmetric under exchange of quantum numbers, with the two fermion constituents occupying different pseudospins.
Notably, the pairing cannot be described as a pure singlet or triplet, because $\bm{k}$ and $\sigma$ transform jointly under the hexagonal-lattice symmetry operations.

\textit{\textbf{Modeling the driven moiré heterostructure.}} --- 
We then investigate how the fermion dynamics discussed above can be implemented using the electronic degrees of freedom in a twisted TMD homobilayer. 
Here the two classes of high-symmetry points [$\sigma = \pm$ in Fig.~\ref{Fig_1}(b)] define two sublattices of the resulting hexagonal superlattice, which are related by an upside-down inversion~\cite{gao2025probing}.
$\sigma$ also serves as the layer pseudospin index, because electronic orbitals near $\sigma = +$ and $-$ are predominantly localized in the top and bottom layers, respectively~\cite{wu2019topological,thompson2025microscopic,zhang2025experimental}. 
We focus on these electronic degrees of freedom in the first valence and conduction moiré bands, with the corresponding transitions induced by detuned, weak external drives arranged in a Raman configuration~\cite{huang2025optical}, together with HPP fluctuations in the surrounding hBN layers.
As these HPPs appear to influence electronic properties~\cite{keren2026cavity}, they can be utilized as a resource to prepare SC in our scenario.

With these considerations, we construct the following model:
\begin{equation}
\label{eq:starting_model}
\hat{H}
=
\hat{H}_{\mathrm{hBN}}
+
\hat{H}_{\mathrm{TMD}}
+
\hat{V}
.
\end{equation}
The first term describes the bare HPP dynamics within hBN, which reads:
\begin{equation}
\hat{H}_{\mathrm{hBN}}
=
\sum_n
\left(\nu_n-\frac{i\gamma_n}{2}\right)
\hat{\alpha}_n^\dagger
\hat{\alpha}_n
,\;
n
\equiv 
(\bm{q}_{||},l)
,
\end{equation}
where $\hat{\alpha}_n$ is the bosonic annilation operator of a HPP with frequency $\nu_n$ and linewidth $\gamma_n\ll \nu_n$ to ensure frequency resolution of these excitations.
$\bm{q}_{||}$ and $l$ label the in-plane momentum and the residual mode index from out-of-plane confinement, respectively.
$\hat{H}_{\mathrm{TMD}}$ and $\hat{V}$ captures the dynamics of the driven moir\'e electrons within the TMD homobilayer and their interaction with HPPs in hBN, respectively.

Specifically, $\hat{H}_{\mathrm{TMD}}$ has the following expression:
\begin{equation}
\hat{H}_{\mathrm{TMD}}
=
\hat{H}_{\mathrm{M}}
+
[
\hat{\mathcal{E}}(t)
+
\hat{\Omega}(t)
+
\mathrm{H.c.}
]
,
\end{equation}
where $\hat{H}_{\mathrm{M}}$ denotes the matter sector with: 
\begin{equation}
\label{eq:H_M}
\hat{H}_{\mathrm{M}}
=
\sum_{\bm{r},\sigma}
(
\omega_c
\hat{c}_{\bm{r},\sigma}^\dagger
\hat{c}_{\bm{r},\sigma}
+
\omega_f
\hat{f}_{\bm{r},\sigma}^\dagger
\hat{f}_{\bm{r},\sigma}
)
+
\hat{V}_{\mathrm{ee}}
,
\end{equation}
where $\hat{c}_{\bm{r},\sigma}$ and $\hat{f}_{\bm{r},\sigma}$ represent the annihilation operators for electrons in the conduction moir\'e band and \textit{holes} in the valence moir\'e band, respectively. 
They play the role of auxiliary and target degrees of freedom mentioned earlier, cf. Eq.~\eqref{eq:jump_general} for the latter.
Here $\omega_c$ and $\omega_f$ denote the corresponding electron and hole energies, which is $\sigma$ independent as the two layers are identical.
For simplicity, we suppress the spin index of these charges, which captures the physics in the presence of an external perpendicular magnetic field.
$\hat{V}_{\mathrm{ee}}$ captures the electrostatic interaction between the electrons and holes, which red-detunes the intralayer (or equivalently, intra-sublattice) transitions while giving only negligible contribution to the interlayer ones due to their differences in the distance between the optically excited electron and hole (see Methods). 

Due to the spectral separation of these transitions, it is possible to drive them independently.
By definition, driving an interlayer transition $\hat{\mathcal{E}}$ flips the layer pseudospin $\sigma$ of the fermions while it is preserved for the corresponding intralayer operation $\hat{\Omega}$ [Fig.~\ref{Fig_1}(b)].
Furthermore, since the layer index is locked to a sublattice, the former transition is accompanied with a spatial transfer to nearest-neighbors, whereas the latter is local.
Their corresponding expressions are (with $\omega \equiv \omega_c+\omega_f$):
\begin{equation}
\label{eq:drive}
\hat{\mathcal{E}}(t)
\simeq
2
\mathcal{E}
e^{-i(\omega-\Delta
)t}
\sum_{\langle\bm{R}_{\bm{r},\sigma},\bm{R}_{\bm{r}', -\sigma}\rangle}
\hat{c}_{\bm{r},\sigma}^\dagger
\hat{f}_{\bm{r}',-\sigma}^\dagger
,
\end{equation}
\begin{equation}
\hat{\Omega}(t)
=
\Omega
e^{-i(\omega-V_B-\Delta - \nu)t}
\sum_{\bm{r},\sigma}
\hat{c}_{\bm{r},\sigma}^\dagger
\hat{f}_{\bm{r},\sigma}^\dagger
,
\end{equation}
with $\mathcal{E}$ and $\Omega$ being the corresponding Rabi frequencies.
$\Delta$ (or $\Delta + \nu$) denotes the detuning of the interlayer (or intralayer) drives from the corresponding excitation gap, which are roughly $\omega
$ (or $\omega - V_B$, with $V_B$ being the binding energy from $\hat{V}_{\mathrm{ee}}$).
$|\Delta|$ thus has to be smaller than $V_B$ and $\nu$ to ensure frequency resolution between these transitions, and meanwhile $|\frac{\nu}{\nu_n}-1|\ll 1$ is chosen such that HPPs can be accessed by the drives.
Notably, the interlayer drives transfer electrons between neighbouring $\sigma = +$ and $-$ supersites, whereas intralayer ones do not.

We model lastly the electron-HPP coupling $\hat{V}$ with the standard Fr\"{o}hlich form where the HPP field couples to the electronic density.
In the dilute limit of conduction electrons, this becomes (see Supplementary Note 1):
\begin{equation}
\label{eq:V}
\hat{V}
=
g
\sum_{\bm{r},\sigma,n}
\sqrt{\frac{\nu_n}{2\epsilon_0}}
[
\Phi_{n,\sigma}(\bm{r})
\hat{\alpha}_n
+
\mathrm{H.c.}
]
\hat{f}_{\bm{r},\sigma}^\dagger
\hat{f}_{\bm{r},\sigma},
\end{equation}
with $g$ capturing the fermion-HPP vertex and $\epsilon_0$ denoting the vacuum permittivity.
Here $\Phi_{n,\sigma}(\bm{r})$ is the HPP mode function integrated over moir\'e-Wannier orbitals of $\hat{f}_{\bm{r},\sigma}$, which has the following expression upon assuming that these orbitals take a Gaussian profile with width $a_W$:
\begin{equation}
\label{eq:Phi_Gaussian}
\Phi_{n,\sigma}(\bm{r})
=
-
\frac{E_n^{||}}{iq_{||}} 
\sqrt{\frac{2\epsilon_0}{\nu_n}}
\exp\left[
i\bm{q}_{||}\cdot\bm{R}_{\bm{r},\sigma}
-
\frac{a_W^2 q_{||}^2}{4}
\right]
.
\end{equation}
Here the exponential term originates from in-plane continuous translational invariance of hBN, and $E_n^{||}$ is the in-plane electric field associated to HPP evaluated at an arbitrary but fixed point at the TMDs.

\textit{\textbf{Effective Lindblad model.}}---
Eq.~\eqref{eq:starting_model} can be simplified to a model expressed in terms of $\hat{f}_{\bm{r},\sigma}$ after perturbative expansions with respect to HPP-fermion coupling, adiabatic elimination of the conduction electrons, and integrating out the HPPs within the Born-Markov approximation.
These treatments are valid provided the hierarchy:
\begin{equation}
\frac{|\mathcal{E}|}{|\Omega|}
\ll
\frac{g\Phi_{n,\sigma}(\bm{r})}{\sqrt{2\epsilon_0\nu_n}}
\ll
\frac{|\mathcal{E}|}{|\Delta|}
,\,
\frac{|\Omega|}{|\Delta|}
,\,
\frac{\gamma_n}{|\Delta|}
\ll
1
,
\end{equation}
and eventually lead to the following Lindblad master equation (see Supplementary Note 2):
\begingroup
\thinmuskip=1.5mu
\medmuskip=2mu
\thickmuskip=2.5mu
\begin{equation}
\label{eq:Lindblad_model}
\begin{aligned}
&
\partial_t
\hat{\rho}
=
\sum_{\bm{r},\bm{r}'}
\sum_{\sigma,\sigma'}
i
\mathcal{J}(\bm{R}_{\bm{r},\sigma}-\bm{R}_{\bm{r}',\sigma})
[
\hat{\rho}
,
\hat{T}_{\bm{r},\sigma}^\dagger
\hat{T}_{\bm{r}',\sigma'}
]
\\
&+
\Upsilon(\bm{R}_{\bm{r},\sigma}-\bm{R}_{\bm{r}',\sigma})
\left[
\hat{T}_{\bm{r}',\sigma'}
\hat{\rho}
\hat{T}_{\bm{r},\sigma}^\dagger
-
\frac{
\{
\hat{\rho}
,
\hat{T}_{\bm{r},\sigma}^\dagger
\hat{T}_{\bm{r}',\sigma'}
\}
}{2}
\right]
,
\end{aligned}
\end{equation}
\endgroup
where the jumps $\hat{T}_{\bm{r},\sigma}$ are defined in Eq.~\eqref{eq:jump_design}, and the couplings $\mathcal{J}(\bm{R})$ and $\Upsilon(\bm{R})$ are set by the driving and dissipation parameters.

The structure for $\hat{T}_{\bm{r},\sigma}$ originates from the amplitudes $u_{\sigma,\sigma'}(\bm{r}-\bm{r}')$ and $v_{\sigma,\sigma'}(\bm{r}-\bm{r}')$ in Eq.~\eqref{eq:design}, which are respectively dictated by dissipation and drive:
$u_{\sigma,\sigma'}(\bm{r}-\bm{r}')$ is effectively local because removing a valence hole (injecting a valence electron) is accompanied by incoherent HPP emission, which carries away which-moiré-site information and thus suppresses spatial coherence.
By contrast, injecting a valence hole (removing a valence electron) is a coherent, drive-induced process [Eq.~\eqref{eq:drive}], so $v_{\sigma,\sigma'}(\bm{r}-\bm{r}')$ can, in principle, couple a given site to multiple neighbors.
In practice, the exponential suppression of wavefunction overlap restricts $v_{\sigma,\sigma'}(\bm{r}-\bm{r}')$ to nearest-neighbor contributions, which in a hexagonal lattice necessarily correspond to inter-sublattice processes.
Together with the rotational symmetry inherited from the drives, this yields
$v_{\sigma,\sigma'}(\bm{r}-\bm{r}') \sim \delta_{\langle \bm{R}_{\bm{r},\sigma},\bm{R}_{\bm{r}',\sigma'} \rangle}$.

The degree of interference between these jumps are encoded in the following coherent and incoherent couplings:
\begin{equation}
\label{eq:J_Upsilon_coeffs}
\begin{bmatrix}
\mathcal{J}({\bm{R}})
\\
\Upsilon({\bm{R}})
\end{bmatrix}
=
\sum_n
\frac{(E_n^{||})^2}{q_{||}^2}
e^{
i\bm{q}_{||}\cdot\bm{R}
-
\frac{a_W^2 q_{||}^2}{2}
}
\begin{bmatrix}
\mathrm{Re}(\mathcal{G}_n)
\\
-2
\mathrm{Im}(\mathcal{G}_n)
\end{bmatrix}
,
\end{equation}
with $\mathcal{G}_n $ being the retarded HPP Green's function: 
\begin{equation}
\mathcal{G}_n 
\equiv 
\frac{
|
2 
\mathcal{E}^\ast 
\Omega
|^2
}{\Delta^2}
\frac{g^2}{\nu_n^2}
\frac{1}{\nu - \nu_n + \frac{i\gamma_n}{2}}
.
\end{equation}
Qualitatively, $\mathcal{J}(\bm{R}-\bm{R}')$ and $\Upsilon(\bm{R}-\bm{R}')$ characterize the HPP field at site $\bm{R}$ emitted from $\bm{R}'$. 
Depending on the spatial structure of these couplings, two limiting regimes can arise: 
In the local-dissipation limit, the emitted modes have sufficient spatial resolution to distinguish different sites, thereby suppressing mutual interference, and giving $\mathcal{J}(\bm{R}-\bm{R}'), \Upsilon(\bm{R}-\bm{R}')\sim \delta_{\bm{R},\bm{R}'}$. 
In the collective regime, by contrast, the emitted modes do not resolve the lattice spacing, and interference between emissions becomes significant. 
In the following sections, we discuss how to achieve these limits respectively within our platform.

\textit{\textbf{Steady state SC with local dissipation.}}---
One strategy to obtain local dissipation from Eq.~\eqref{eq:J_Upsilon_coeffs} is to make the oscillatory term $e^{i\bm{q}_{||}\cdot\bm{R}}$ dominant over all other $\bm{q}_{||}$ dependence in the summand.
The key requirement for realizing this condition is that the bosonic excitations mediating dissipation be effectively nondispersive (irrespective of whether they are hyperbolic).
For HPPs, this can be achieved by restricting to a single $l$ mode over a momentum range larger than $\frac{2\pi}{a_M}$, provided that the corresponding segment exhibits suppressed $\bm{q}_{||}$ dependence in both its dispersion and mode function.

Fig.~\ref{Fig_2}(a) shows that the $l = 0$ mode at momenta $q_{||}\gtrsim\frac{2\pi}{d}$, where $d$ is the hBN thickness, satisfies these criteria.
At the implementation level, these relevant modes can be selected by tuning $\nu$ close to $\nu_{\mathrm{LO}}$.
Specifically, all modes with small $q_{||}$ are filtered out by the small linewidth $\gamma_n$.
Otherwise momenta are summed, up to $q_{||}\sim\frac{2\pi}{a_W}$ due to the ultraviolet cutoff $e^{-\frac{a_W^2 q_{||}^2}{2}}$ in Eq.~\eqref{eq:J_Upsilon_coeffs}, originating from the width of electronic orbitals $a_W$.
Note that the remaining wide $q_{||}$ window condition further suggests $\frac{2\pi}{a_W} - \frac{2\pi}{d} \gg \frac{2\pi}{a_M}$, which can be fulfilled in hBN-encapsulated moir\'e TMDs --- typically $a_W\sim 2$nm and $a_M\sim10$nm, and $d$ is tunable between $\sim 1 - 100$ nm.
These expectations are consistent with numerical results [Fig.~\ref{Fig_2}(b)]: $\Upsilon(\bm{R}) \sim \delta_{\bm{R},0}$ for large values of $\frac{a_M}{a_W}$, and longer-ranged contributions grows at smaller values of such ratio (see Supplementary Note~3).

Notably, in addition to these Lindblad terms, there is a Lamb shift Hamiltonian $\hat{\mathcal{H}} = J\sum_{\bm{r},\sigma} \hat{T}_{\bm{r},\sigma}^\dagger \hat{T}_{\bm{r},\sigma}$.
It nevertheless has the dark state as an eigenstate, and therefore, does not compete with dissipative evolution as already noted.

\begin{figure}
\centering
\includegraphics[width=\linewidth]{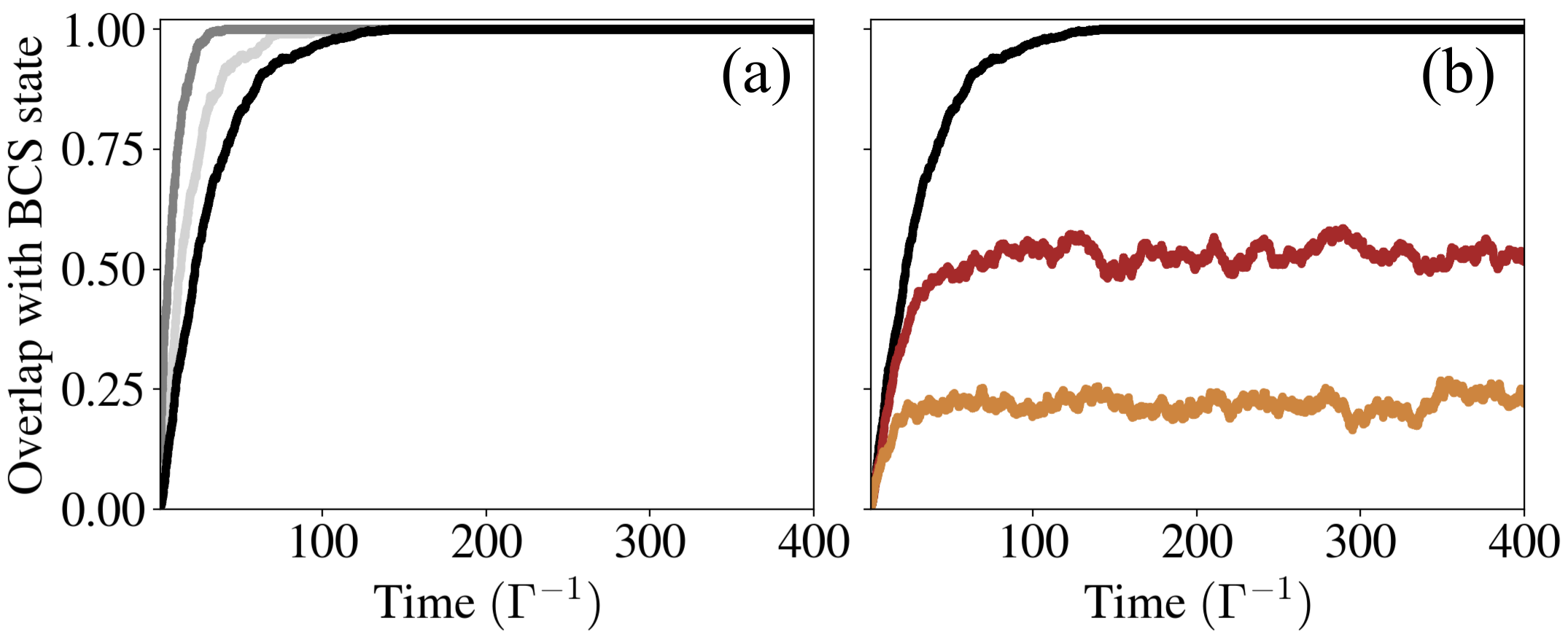}
\caption{
\textbf{Emergence of superconductivity upon time evolution}.
The overlap of the time-evolved density matrix with the BCS state $|\text{BCS},N\rangle$ is computed under various circumstances.
To this end we numerically solve  Eq.~\eqref{eq:Lindblad_model_SC} using the quantum-trajectory method with a total of $500$ trajectories.
(a) Different fermion fillings.
Light gray, gray, and black points correspond to systems with $2$, $4$, and $6$ fermions, respectively, on a $12$-site hexagonal lattice with periodic boundary conditions.
Initial states are chosen such that all fermions occupy a single sublattice.
(b) Different degrees of imperfection $p$, involved in the jump as $\hat{T}_{\bm{r},\sigma}\to\hat{T}_{\bm{r},\sigma} + p \hat{f}_{\bm{r},\sigma}^\dagger \hat{f}_{\bm{r},\sigma}$.
Black, brown, and light brown represents $p = 0$, $0.1$, and $0.2$, respectively.
All other conditions are the same as the black curve in (a).
Time is displayed in units of dissipation $\Upsilon_{\sigma,\sigma}(0) \equiv \Gamma$.
}
\label{Fig_3}
\end{figure}

The emergence of the BCS state upon time evolution can be numerically confirmed, using quantum trajectory method~\cite{daley2014quantum}, for small system size over a range of fermion fillings [Fig.~\ref{Fig_3}(a)].
The convergence further suggests uniqueness of the stationary state, which can be further benchmarked with larger system size with a complementary approach [see Methods].
Even in the presence of imperfection ($p$), we expect this overlap, and hence our scheme, to be robust in the thermodynamic limit as long as $p = \Delta \Gamma/\Gamma \ll 1$, where $\Delta \Gamma$ is the rate or the energy scale of perturbation: 
A mean field decoupling of dark state Lindbladians of the type of Eq.~\eqref{eq:Lindblad_model} utilizing the proximity to the exactly known dark state shows that the unperturbed dynamics features a dissipative gap $\sim \Gamma$, setting a scale of robustness against perturbations in analogy to an energy gap in equilibrium systems~\cite{bardyn2013njp}.
To demonstrate this, we exemplarily consider $\hat{T}_{\bm{r},\sigma}\to\hat{T}_{\bm{r},\sigma} + p \hat{f}_{\bm{r},\sigma}^\dagger \hat{f}_{\bm{r},\sigma}$ in finite size numerics, where the extra term can be incorporated by including the intralayer Raman transition (dropped earlier as it is far detuned).
The corresponding stationary state still possesses a nontrivial overlap with the BCS state if $p$ is perturbative [Fig.~\ref{Fig_3}(b)].

\begin{figure}
\centering
\includegraphics[width=\linewidth]{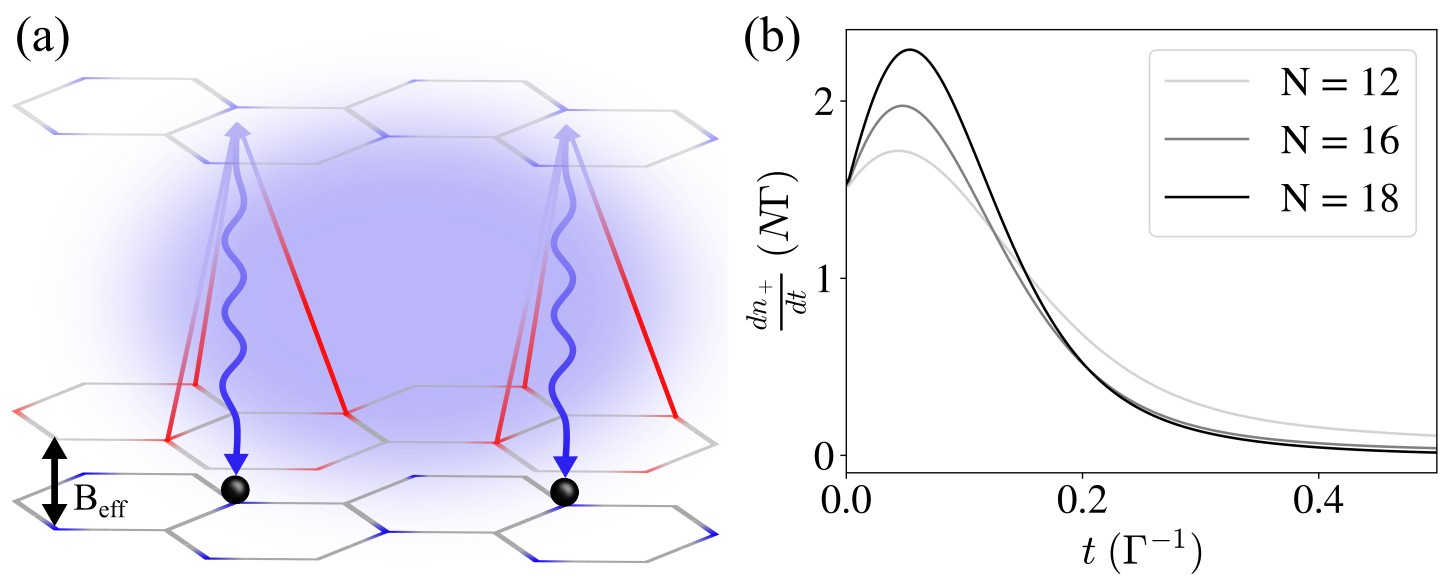}
\caption{
\textbf{Illustration of collective dissipation and its consequence.}
(a) In contrast to Fig.~\ref{Fig_1}(b), here the dissipators are nonlocal, thereby interfering with each other (indicated by blurred region), and acting collectively.
Upon lifting the degeneracy in layer pseudospin with an effective Zeeman field $B_{\text{eff}}$, the collective dissipation leads to a superradiant burst in the pseudospin transition rate $\frac{dn_+}{dt}$ as demonstrated in (b).
Results in this panel are numerically performed for a range of system sizes $N$, by directly evaluating the Lindblad equation~\eqref{eq:Lindblad_model_SR} starting from an initial state where all $\sigma = -$ sites are filled.
}
\label{Fig_4}
\end{figure}

\textit{\textbf{Transient superradiance with collective dissipation.}}---
In contrast, the uniqueness of the dark state enforced by Eq.~\eqref{eq:dark_state_condition} no longer holds in situations where dissipation is nonlocal: $\Upsilon(\bm{R}-\bm{R}')\neq 0,\quad\forall \bm{R}\neq\bm{R}'$.
This is because dissipation across different sites can destructively interfere [Fig.~\ref{Fig_4}(a)], which quenches the corresponding dissipative channels, thereby generating extra dark states.
Although the absence of dark state uniqueness suggests that preparation of SC is not ideal with nonlocal dissipation, such a regime still gives rise to interesting transient physics, as we now elaborate.

We consider the extreme limit of fully collective dissipation, $\Upsilon(\bm{R}_{\bm{r},\sigma}-\bm{R}_{\bm{r}',\sigma'}) = w_\sigma \Gamma w_{\sigma'}$, where the dissipative coupling is uniformly all-to-all except for a pseudospin-dependent weighting factor $w_\sigma$. 
Together with a similar form for the Hamiltonian dynamics $\mathcal{J}(\bm{R}_{\bm{r},\sigma}-\bm{R}_{\bm{r}',\sigma'}) = w_\sigma J w_{\sigma'}$,
Eq.~\eqref{eq:Lindblad_model} can be recasted as:
\begin{equation}
\label{eq:Lindblad_model_SR}
\partial_t\hat{\rho}
=
-iJ
[
\hat{T}^\dagger
\hat{T}
,
\hat{\rho}
]
+
\Gamma
\left(
\hat{T}
\hat{\rho}
\hat{T}^\dagger
-
\frac{1}{2}
\{
\hat{T}^\dagger
\hat{T}
,
\hat{\rho}
\}
\right)
,
\end{equation}
where the \textit{only} decay channel is given by the following jump operator:
\begin{equation}
\hat{T}
=
\sum_{\bm{k}}
w_-
\eta_{\bm{k}}^\ast
\hat{\mathcal{S}}_{\bm{k}}^{-}
+
w_+
\eta_{\bm{k}}
\hat{\mathcal{S}}_{\bm{k}}^{+}
,\quad
\eta_{\bm{k}}
=
\sum_{\bm{\delta}}
e^{i\bm{k}\cdot\bm{\delta}}
,
\end{equation}
where $\{\bm{\delta}\}$ are the set of vectors pointing from a $\sigma = -$ site to its nearest neighbors.
$\hat{\mathcal{S}}_{\bm{k}}^{\pm}=\check{f}_{\bm{k}}^\dagger\cdot \check{\sigma}^{\pm}\cdot\check{f}_{\bm{k}}$ is the raising/lowering operator for pseudospins defined at each $\bm{k}$.
Crucially, because there is now only a single collective jump operator $\hat{T}$, the dark subspace is no longer unique: the \textit{only} constraint is $\hat{T}| D\rangle = 0$, for which multiple solutions can exist.

Notably, $\hat{T}$ plays the role of a collective spin operator defined on a ($\bm{k}$-space) array. 
In the limit $w_+/w_- \to 0$, this setting closely resembles arrays of two-level optical emitters~\cite{masson2022universality,huang2025collective,kumlin2025superradiance,lagnese2026neural}, with the difference that those arrays are usually considered in real space. 
A characteristic many-body phenomenon in such systems is that, starting from an initial state in which the pseudospins at all sites are in the excited flavor, the spin-flip (or emission) rate exhibits a short-time (superradiant) burst.
The emergence of a burst in the sublattice transition rate is numerically confirmed when all $\sigma = -$ sites are initially filled [Fig.~\ref{Fig_4}(b)], neglecting the Hamiltonian dynamics~\cite{masson2022universality,kumlin2025superradiance}, which does not affect the presence of the burst (see Supplementary Note~4).

The switching from local to collective dissipation can be achieved in our moiré heterostructure platform by simply turning off the $\Omega$ drive. 
In this case, the $\mathcal{E}$ drive yields dissipation associated with radiation at frequencies near the band gap, rather than coupling to the HPPs.
The corresponding dissipation matrix $\Upsilon(\bm{R}-\bm{R}')$ is then governed by the (imaginary part of the) free-space electromagnetic Green's function~\cite{huang2025collective,lagnese2026neural}, which, in our platform where the superlattice spacing is much smaller than the optical wavelength, is approximately all-to-all (at least for finite superlattices). 
Meanwhile, the sublattice imbalance $w_\pm$ can be introduced via an out-of-plane electric field that shifts the energies of the two TMD layers, which acts as a Zeeman splitting between the pseudospins. 
Specifically, this frequency resolution enables selective driving of the interlayer transition $\sigma = - \to +$ without $\sigma = + \to -$, effectively realizing the limit $w_+/w_- \to 0$, thereby implementing the conditions under which the burst in Fig.~\ref{Fig_4}(b) appears.

\textit{\textbf{Discussion and Outlook.}} --- 
We anticipate that several aspects of the dissipatively prepared steady state SC can be probed via techniques applied for equilibrium states.
In particular, such a generalizability is expected when merely the properties of the state --- but not its finite-frequency dynamics --- are probed.
This includes, for example, HPP emission intensity and optical conductivity, which respectively signal the emergence of stationarity and the  superfluid stiffness.

In addition to our nonequilibrium order, intrinsic superconductivity in moiré TMD homobilayers may also contribute to these responses. 
These undriven orders encode intriguing physics: 
in tWSe\textsubscript{2}~\cite{xia2025superconductivity,guo2025superconductivity}, SC appears near Mott insulating regimes and has therefore been speculated to arise from antiferromagnetic fluctuations~\cite{qin2025topological}, resembling the paradigmatic spin-mediated mechanism toward high-temperature superconductivity~\cite{chubukov2008spin}; 
in tMoTe\textsubscript{2}~\cite{xu2025signatures}, it can even emerge together with fractional quantum anomalous Hall physics, further motivating possible routes toward anyon superconductivity~\cite{divic2025anyon,shi2026superconductivity}. 
These intriguing orders could be reproduced or further stabilized within our scenario through suitably adapted engineered drive and dissipation, thereby offering a nonequilibrium route toward realizing and controlling these unconventional superconducting states.


Finally, the driven-dissipative nature of our preparation protocol has the potential to open a new arena for non-equilibrium statistical mechanics in the quantum domain. 
Breaking microscopic equilibrium conditions alone can already generate phases of matter with universal long-distance properties that have no equilibrium counterpart. 
A striking example is provided by recent observations of nonequilibrium Kardar-Parisi-Zhang scaling in the Goldstone-mode dynamics of driven polariton condensates~\cite{Fontaine2022Nature,Widmann2026Science}. 
Light-induced superconductors similarly realize intrinsically nonequilibrium states, while enriching the problem through conservation laws and the presence of  electromagnetic fields, both of which can shape the universal long-wavelength behavior via hydrodynamic modes and gapless gauge fluctuations~\cite{Sentef_2016,Eckhardt2024PhotoSC,Kennes_2017,DiesselSachdevBonetti2026}, 
A common feature of these platforms is the presence of finite noise, which ultimately drives the system into a mixed quantum state and, at the longest scales, classical nonequilibrium dynamics. 
Our setting combines the same essential ingredients --- spontaneous symmetry breaking, conservation laws, and gauge fields --- but differs in one crucial aspect: the dissipatively prepared steady state is pure, providing a nonequilibrium analogue of a zero-temperature equilibrium state. 
This opens the door to a genuinely \textit{quantum} nonequilibrium statistical mechanics, in which many-body quantum dynamics and universal nonequilibrium scaling coexist, and whose phenomenology remains largely unexplored.

\bibliography{Biblio}

\clearpage

\section{Methods}

\subsection{Properties of HPPs in hBN}

In this section, we review the properties of HPPs in hBN, following Ref.~\cite{feng2026color}.
To begin with, we consider the hBNs as a slab of nonmagnetic medium with dielectric tensor $\text{diag}(\epsilon_{\parallel},\epsilon_{\parallel},\epsilon_{\perp})$, with $\parallel$ and $\perp$ denoting the (two) in-plane and (one) out-of-plane directions, respectively.
The transverse-magnetic (TM) modes therein satisfy the following generic dispersion relation:
\begin{equation}
\frac{q^2}{\epsilon_{\perp}(\nu)}
+
\frac{q_{\perp}^2}{\epsilon_{\parallel}(\nu)}
=
\frac{\nu^2}{c^2}
,
\end{equation}
where $q_{\perp}$ and $q\equiv q_{\parallel}$ denote components of the HPP momentum, and $c$ is the speed of light.
Hyperbolicity is this dispersion emerges when $\epsilon_{\perp}\epsilon_{\parallel}<0$, which occurs in hBN due to the electromagnetic resonances from anisotropic phonons therein (HPPs).
In particular, in the deep subwavelength limit where $c\sqrt{q^2+q_{\perp}^2}\gg \nu$, this leads to:
\begin{equation}
\label{eq:kappa}
\kappa
\equiv
\frac{q}{q_\perp}
\simeq
\sqrt{-\frac{\epsilon^{\perp}(\nu)}{\epsilon^{\parallel}(\nu)}}
,
\end{equation}
which sets of angle of HPP propagation.
That is, such a direction is determined by that between dielectric constants, which are generally taken to be the following form~\cite{su2024fundamentals}:
\begin{equation}
\label{eq:hBN_dielec}
\epsilon_{\parallel}(\nu)
\simeq
\epsilon_{\parallel}(\infty)
\frac{
\nu_{\parallel,\mathrm{LO}}^2-\nu^2
}{
\nu_{\parallel,\mathrm{TO}}^2-\nu^2
}
,
\end{equation}
and similarly for $\epsilon_{\perp}(\nu)$.
Here, $\nu_{\parallel,\mathrm{LO}}$ and $\nu_{\parallel,\mathrm{TO}}$ ($\nu_{\perp,\mathrm{LO}}$ and $\nu_{\perp,\mathrm{TO}}$) are frequencies of longitudinal and transversal optical phonons along the in-plane (out-of-plane) direction, and $\gamma_{\parallel}$ ($\gamma_{\perp}$) denotes the corresponding loss.

In additional to Eq.~\eqref{eq:kappa}, another constraint between the momentum components is given by the quantization condition of the slab with out-of-plane width $d$.
Specifically, utilizing the standard electromagnetic boundary conditions to match the electric fields inside and outside the slab, one obtains:
\begin{equation}
\label{eq:HPP_disp}
q
= 
\frac{\kappa}{d}
\left[
2
\tan^{-1}\left(-\frac{1}{\epsilon_{\parallel}\kappa}\right)
+
l\pi
\right]
,
\end{equation}
where $l$ is the mode index originating from this out-of-plane confinement.

Here we specifically focus on the frequency range $\nu_{\parallel,\mathrm{TO}}\leq\nu\leq\nu_{\parallel,\mathrm{LO}}$, where $\epsilon_{\parallel}(\nu)<0$ and $\epsilon_{\perp}(\nu)>0$.
Combining Eq.~\eqref{eq:kappa}, Eq.~\eqref{eq:hBN_dielec}, and Eq.~\eqref{eq:HPP_disp}, and utilizing the numbers in Extended Data Table~\ref{tab:HPP_paras} (with the losses $\gamma_{\parallel}$ and $\gamma_{\perp}$ dropped), we arrive at the dispersion relation in Fig.~\ref{Fig_2}(a).

In addition, we note that the HPP loss rate, $\gamma \equiv \gamma_{\parallel}$, is chosen in an optimal regime: 
On the one hand, $\gamma$ is sufficiently large that the suppressed HPP group velocity prevents excitations from propagating to neighboring sites and validates the Born-Markov approximation. 
On the other hand, $\gamma$ is sufficiently small that higher-order modes $l>0$ remain outside the linewidth and therefore contribute negligibly compared with the $l=0$ mode, obeying the condition described in the main text.

\subsection{Momentum space representation}

In this section, we elaborate on the momentum space representations of operators and quantities in the hexagonal lattice and the derivation of relevant relations.
To begin with, the Fourier transformation of the target fermions is defined as:
\begin{equation}
\label{eq:fermion_k}
\hat{f}_{\bm{k},\sigma}
=
\sqrt{\frac{2}{N}} 
\sum_{\bm{r}} 
e^{-i\bm{k}\cdot\bm{R}_{\sigma}(\bm{r})} 
\hat{f}_{\bm{r},\sigma}
,
\end{equation}
with $N$ ($N/2$) being the number of sites (unit cells), and $\bm{R}_{\sigma}(\bm{r})$ being the position vector of the sites, which, in a hexagonal lattice with periodicity $a_M$, reads:
\begin{equation}
\bm{R}_{\sigma}(\bm{r})
=
\bm{r}
+
\text{sgn}_{\sigma}
\frac{a_M}{\sqrt{3}}
\bm{e}_y
,
\end{equation}
where $\text{sgn}_A =-\text{sgn}_B= 1$, and:
\begin{equation}
\frac{\bm{r}}{a_M}
=
\left(
n_1
+
\frac{n_2}{2}
\right)
\bm{e}_x
+
\frac{\sqrt{3}n_2}{2}
\bm{e}_y
,
\end{equation}
with integers $n_1$ and $n_2$ labeling the unit cells.
In the spirit of Eq.~\eqref{eq:fermion_k}, we define the creators and annihilators in the Bogoliubov eigenoperators as:
\begin{equation}
\begin{aligned}
\begin{bmatrix}
\hat C_{\bm{k},\sigma}^\dagger
\\
\hat A_{\bm{k},\sigma}
\end{bmatrix}
&
\equiv
\sqrt{\frac{2}{N}} 
\sum_{\bm{r}} 
\begin{bmatrix}
e^{i\bm{k}\cdot\bm{R}_{\sigma}(\bm{r})} 
\hat C_{\bm{r},\sigma}^\dagger
\\
e^{-i\bm{k}\cdot\bm{R}_{\sigma}(\bm{r})} 
\hat A_{\bm{r},\sigma}
\end{bmatrix}
\\
&
=
\sum_{\sigma'}
\begin{bmatrix}
v_{\sigma,\sigma'}(\bm{k}) \hat f_{\bm{k},\sigma'}^\dagger
\\
u_{\sigma,\sigma'}(\bm{k}) \hat f_{\bm{k},\sigma'}
\end{bmatrix}
,
\end{aligned}
\end{equation}
where we defined:
\begin{equation}
\begin{bmatrix}
v_{\sigma,\sigma'}(\bm{k})
\\
u_{\sigma,\sigma'}(\bm{k})
\end{bmatrix}
\equiv
\sum_{\bm{r}}
\begin{bmatrix}
e^{i\bm{k}\cdot[\bm{R}_\sigma(\bm{r})-\bm{R}_{\sigma'}(0)]} 
v_{\sigma,\sigma'}(\bm{r})
\\
e^{-i\bm{k}\cdot[\bm{R}_\sigma(\bm{r})-\bm{R}_{\sigma'}(0)]} 
u_{\sigma,\sigma'}(\bm{r})
\end{bmatrix}
.
\end{equation}

\subsection{Implementation with symmetric drive}

A direct implementation of drives that realize $\check{v}(\bm{r})$ in Eq.~\eqref{eq:design} would require anisotropy in the layer-sublattice pseudospin, adding an extra layer of complexity. 
To avoid it, we instead consider drives that generate $\check{\bar v}(\bm{r}) = \check{\sigma}^z \check{v}(\bm{r})$. 
This alternative choice is valid because the replacement $\check{v}(\bm{r}) \to \check{\bar v}(\bm{r})$ merely changes the jump operator as $\hat T_{\bm r,\sigma} \to \sigma \hat T_{\bm r,\sigma}$; this modification is inconsequential for the Lindbladian in Eq.~\eqref{eq:Lindblad_model_SC} and therefore preserves the dark state.

\subsection{Uniqueness of dark state}

One direct way to probe the uniqueness of the dark state is to compute the time evolution from a set of randomly chosen initial states and verify that the long-time state converges to a pure steady state with unit overlap with the target (here the BCS state), as done for Fig.~\ref{Fig_2}(b). 
This approach is, however, numerically expensive because it requires repeated long-time propagation for many initial conditions. 
Here we instead assess uniqueness by analysing the eigenspectrum of the emergent operator $\sum_{\bm{r},\sigma} \hat{T}_{\bm{r},\sigma}^\dagger \hat{T}_{\bm{r},\sigma}$: 
Its eigenstate at zero eigenvalue is the dark state, since $\hat{T}_{\bm{r},\sigma}|D\rangle = 0$. 
Consequently, a non-degenerate zero eigenvalue implies that $|D\rangle$ is unique. 
Notably, this method requires only the lowest few eigenstates of a single matrix, and is therefore limited primarily by memory through the matrix dimension, which is determined by the fermion number and lattice size. 
In Extended Data Fig.~\ref{ext_Fig_1}, we fix either of these numbers and plot the low-lying spectrum of $\sum_{\bm{r},\sigma} \hat{T}_{\bm{r},\sigma}^\dagger \hat{T}_{\bm{r},\sigma}$ while varying the other. 
We find that, at least for small filling fractions on small lattices, the lowest two eigenstates remain gapped, suggesting a unique dark state in this regime.

\subsection{Electron-electron interaction}

Regarding electron-electron interaction, we specifically consider only the on-site interband electron-hole attraction:
\begin{equation}
\hat{V}_{\mathrm{ee}}
= 
- V_B 
\sum_{\bm{r},\sigma}  
\hat{c}_{\bm{r},\sigma}^{\dagger} 
\hat{f}_{\bm{r},\sigma}^{\dagger} 
\hat{f}_{\bm{r},\sigma}  
\hat{c}_{\bm{r},\sigma}  
,
\end{equation}
with $V_B>0$ capturing the strength of such attraction.
This is because the repulsion between valence electrons vanish due to Pauli exclusion, and similar term between (dilute) conduction electrons is negligible as we focus on the weak drive limit.
In addition, we gate the bilayer by metallic plates nearby to suppress all off-site interactions.

\subsection{Use of AI tools}

During preparation of this manuscript, the authors used ChatGPT to assist with language polishing, improving codes, and double checking calculations. 
The tool was not used to generate research data, perform the scientific analysis, or draw conclusions. 
All AI-assisted output was reviewed, edited and verified by the authors, who take full responsibility for the content of the manuscript.

\clearpage

\section{Data availability}
All data supporting the findings of this study are included in the main text and the Supplementary Information. 
Source data are available from the corresponding authors upon request.

\section{Acknowledgements}
T.-S.H. thanks D. E. Chang for useful discussions and acknowledges support from the European Union, under European Research Council grant agreement No 101002107 (NEWSPIN); the Government of Spain (Severo Ochoa Grant CEX2024-001490-S [MICIU/AEI/10.13039/501100011033]); Generalitat de Catalunya (CERCA program); Fundació Cellex, and Fundació Mir-Puig.
T.-S.H. and M.H. were supported by AFOSR MURI FA9550- 19-1-0399 and Simons Foundation. 
A.I. was supported by the Swiss National Science Foundation (SNSF) under Grant No. 2000-1-240035.
S.D. was supported by the Deutsche Forschungsgemeinschaft (DFG, German Research Foundation) under Germany’s Excellence Strategy Cluster of Excellence Matter and
Light for Quantum Computing (ML4Q) EXC 2004/1390534769, and CRC 1238 project number 277146847.
This research was supported in part by grant NSF PHY-2309135 to the Kavli Institute for Theoretical Physics (KITP).

\section{Competing interests}
The authors declare no competing interests.

\section{Contributions}
All authors conceived the project. 
T.-S. H. did the theoretical calculation, and A. I., S. D., and M. H. contributed to the theoretical understanding. 
All authors contributed the writing of the manuscript, and A. I., S. D., and M. H. supervised the work. 
All authors discussed the results and contributed to the manuscript.

\clearpage

\onecolumngrid

\section{Extended data figures and tables}
\setcounter{figure}{0}  

\begin{table}[h]
\begingroup
\renewcommand{\tablename}{\textbf{Extended Data Table}}
\renewcommand{\thetable}{\arabic{table}}
\renewcommand{\theHtable}{ED.\arabic{table}}
\centering
\begin{tabular}{|c|c|c|c|c|c|c|}
\hline
$\epsilon_{\parallel}(\infty)$ 
&
$\epsilon_{\perp}(\infty)$
&
$\nu_{\parallel,\mathrm{LO}}$
&
$\nu_{\parallel,\mathrm{TO}}$
&
$\nu_{\perp,\mathrm{LO}}$
&
$\nu_{\perp,\mathrm{TO}}$
&
$\gamma_n$
\\
\hline
$4.9$
&
$2.95$
&
$48.42$
& 
$40.8$
&  
$24.75$
& 
$22.8$
&
$0.13$
\\
\hline
\end{tabular}
\caption{
\textbf{Parameters of hexagonal boron nitride.}
Dielectric constants ($\epsilon_{\parallel}(\infty)$ and $\epsilon_{\perp}(\infty)$) and phonon frequencies ($\nu_{\parallel,\mathrm{LO}}$, $\nu_{\parallel,\mathrm{TO}}$, $\nu_{\perp,\mathrm{LO}}$, and $\nu_{\perp,\mathrm{TO}}$) are from Ref.~\cite{feng2026color}, and decay rate ($\gamma$) from Refs.~\cite{giles2018ultralow,herzig2024high}.
Frequencies and decay rate are in units of THz.
}
\label{tab:HPP_paras}
\endgroup
\end{table}

\begin{figure}[h]
\centering
\begingroup
\renewcommand{\figurename}{\textbf{Extended Data Fig.}}%
\renewcommand{\thefigure}{\arabic{figure}}
\renewcommand{\theHfigure}{ED.\arabic{figure}}
\includegraphics[width=0.6\linewidth]{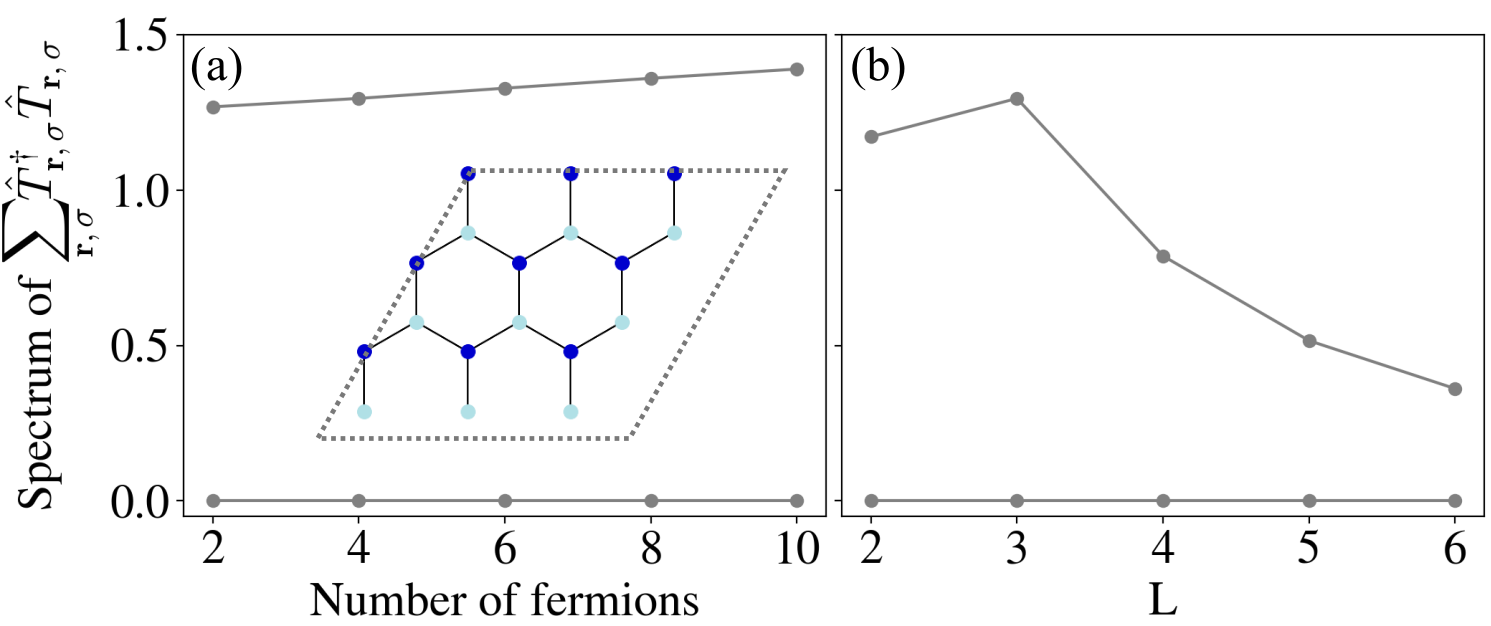}
\caption{
\textbf{Uniqueness of the dark state}.
Uniqueness is diagnosed from the zero-eigenvalue sector of the emergent operator $\sum_{\bm{r},\sigma} \hat{T}_{\bm{r},\sigma}^\dagger \hat{T}_{\bm{r},\sigma}$: a non-degenerate ground state implies a unique dark state.
(a) Low-lying spectrum for different fermion numbers on an $18$-site hexagonal lattice (inset).
(b) Same as (a), but with the fermion number fixed to $4$ and the lattice size varied, with the number of sites $N_s = 2L^2$.
}
\label{ext_Fig_1}
\endgroup
\end{figure}

\end{document}


\begin{CJK}{UTF8}{gbsn}

\title{Driven-dissipative superconductivity in moiré heterostructure without attraction}

\author{Tsung-Sheng Huang}
\affiliation{Joint Quantum Institute, University of Maryland, College Park, MD 20742, USA}
\affiliation{ICFO-Institut de Ciencies Fotoniques, The Barcelona Institute of Science and Technology, 08860 Castelldefels (Barcelona), Spain}

\author{Ataç Imamoglu}
\affiliation{Institute of Quantum Electronics, ETH Zürich, 8093 Zürich, Switzerland}

\author{Mohammad Hafezi}
\affiliation{Joint Quantum Institute, University of Maryland, College Park, MD 20742, USA}

\author{Sebastian Diehl}
\affiliation{Institute for Theoretical Physics, University of Cologne, 50937 Cologne, Germany}

\date{\today}

\maketitle
\end{CJK}

\section*{Supplementary Note 1 \textbar{} Coupling between electrons and hyperbolic phonon polaritons (HPPs)}
In this section, we justify the following expression for electron-HPP interaction:
\begin{equation}
\label{eq:V}
\hat{V}
=
g
\sum_{\bm{r},\sigma,n}
\sqrt{\frac{\nu_n}{2\epsilon_0}}
[
\Phi_{n,\sigma}(\bm{r})
\hat{\alpha}_n
+
\mathrm{H.c.}
]
\hat{f}_{\bm{r},\sigma}^\dagger
\hat{f}_{\bm{r},\sigma}
.
\end{equation}
To reiterate, here $\hat{f}_{\bm{r},\sigma}$ is the annihilation operator for target fermions --- the \textit{holes} in the first valence moiré band in the transition metal dichalcogenide (TMD) bilayer --- at unit cell $\bm{r}$ and sublattice $\sigma$.
$\hat{\alpha}_n$ is the bosonic annihilator for HPPs, where $n\in(\bm{q}_\parallel,l)$ collects the in-plane momentum $\bm{q}_\parallel$ and mode index $l$, and $\nu_n$ denotes the corresponding HPP frequency.
$g$ captures the coupling strength and $\epsilon_0$ is the vacuum susceptibility.
$\Phi_{n,\sigma}(\bm{r})$ is the HPP mode function integrated over moir\'e-Wannier orbitals of $\hat{f}_{\bm{r},\sigma}$, which we elaborate in the derivation below.

We begin with the Fr\"{o}hlich-type coupling:
\begin{equation}
\label{eq:Frohlich}
\hat{V}_{\mathrm{Fr}}
=
-
g
\int
d^2\bm{x}
\hat{\phi}(\bm{x})
\hat{\psi}^\dagger(\bm{x})
\hat{\psi}(\bm{x})
,
\end{equation}
where $\hat{\psi}(\bm{x})$ denotes the electron field operator evaluated at in-plane position $\bm{x}$. 
Qualitatively, such an interaction captures how the electron density leads to local distortion, here associated to HPPs.
The corresponding scalar field is captured by $\hat{\phi}(\bm{x})$, which, upon standard mode expansion, reads:
\begin{equation}
\hat{\phi}(\bm{x})
=
\sum_{n}
\sqrt{\frac{\nu_n}{2\epsilon_0}}
[
\phi_n(\bm{x})
\hat{\alpha}_n
+
\phi_n^\ast(\bm{x})
\hat{\alpha}_n^\dagger
]
,
\end{equation}
where $\phi_n(\bm{x})$ is the mode function of the scalar potential evaluated at the two dimensional plane.

The next move is to express the electron field operators in the moir\'e-Wannier basis.
Rigorously speaking, in this expansion one should include the electronic states from all moiré bands; nevertheless, since we assume the system is weakly driven, all other states except the ones in the first conduction and valence moiré bands contribute negligibly. 
Thus, we can express:
\begin{equation}
\hat{\psi}(\bm{x})
\simeq 
\sum_{\bm{r},\sigma}
W_c(\bm{x}-\bm{R}_{\bm{r},\sigma})
\hat{c}_{\bm{r},\sigma}
+
W_f(\bm{x}-\bm{R}_{\bm{r},\sigma})
\hat{f}_{\bm{r},\sigma}^\dagger
,
\end{equation}
where $\hat{c}_{\bm{r},\sigma}$ is the annihilation operator for the auxiliary fermions --- electrons in the first conduction moiré band of the TMD bilayer --- with $W_c(\bm{x}-\bm{R}_{\bm{r},\sigma})$ denoting the corresponding moiré-Wannier orbitals at site $\bm{R}_{\bm{r},\sigma}$ ($W_f(\bm{x}-\bm{R}_{\bm{r},\sigma})$ defined similarly for $\hat{f}_{\bm{r},\sigma}$).
The resulting density operator $\hat{\psi}^\dagger(\bm{x})\hat{\psi}(\bm{x})$ involves three terms $\sim \hat{c}_{\bm{r},\sigma}^\dagger\hat{c}_{\bm{r}',\sigma'}$, $\sim \hat{c}_{\bm{r},\sigma}\hat{f}_{\bm{r}',\sigma'}$, and $\sim \hat{f}_{\bm{r},\sigma}\hat{f}_{\bm{r}',\sigma'}^\dagger$.
We drop the first contribution as the conduction band is only virtually accessed.
The second one is also dropped, because the corresponding term in $\hat{V}_{\mathrm{Fr}}$ evolves (in the interaction picture) as $\sim e^{-i(\omega\pm \nu_n)t}$, which is washed out over time because the bandgap $\omega\sim 1.5$eV is much larger than the HPPs frequency $\sim 0.1$eV.
Hence, we keep only the last contribution, which yields:
\begin{equation}
\begin{aligned}
\hat{\psi}^\dagger(\bm{x})
\hat{\psi}(\bm{x})
\simeq
\sum_{\bm{r},\sigma}
|W_f(\bm{x}-\bm{R}_{\bm{r},\sigma})|^2
\hat{f}_{\bm{r},\sigma}
\hat{f}_{\bm{r},\sigma}^\dagger
,
\end{aligned}
\end{equation}
where we assume that these orbitals are much narrower than the moir\'e period such that the off-site contributions are negligible.
Putting this back, we find $\hat{V}_{\mathrm{Fr}}
\simeq V$ in Eq.~\eqref{eq:V} upon normal ordering with respective to the hole operators, with the extra generated term absorbed into the definition of $\hat{\alpha}_n$.
The integrated mode function $\Phi_{n,\sigma}(\bm{r})$ follows as:
\begin{equation}
\label{eq:phi_W}
\Phi_{n,\sigma}(\bm{r})
=
\int
d^2\bm{x}
\phi_n(\bm{x})
|W_f(\bm{x}-\bm{R}_{\bm{r},\sigma})|^2
.
\end{equation}

\section*{Supplementary Note 2 \textbar{} Derivation of the master equation from model for the moir\'e hreterostructure.}

In this Appendix, we discuss the details in derivation of the Lindblad master equation outlined in the main text, which is reproduced below:
\begin{equation}
\label{eq:Lindblad_eqn_reprod}
\partial_t 
\hat{\rho}
= 
\sum_{\bm{r},\bm{r}'} 
\sum_{\sigma,\sigma'}
i 
\mathcal{J}(\bm{R}_{\bm{r},\sigma}-\bm{R}_{\bm{r}',\sigma'})
[
\hat{\rho}
, 
\hat{T}^{\dagger}_{\bm{r},\sigma} 
\hat{T}_{\bm{r}',\sigma'}
]
+ 
\Upsilon(\bm{R}_{\bm{r},\sigma}-\bm{R}_{\bm{r}',\sigma'})
\left[
\hat{T}_{\bm{r}',\sigma'} 
\hat{\rho} 
\hat{T}^{\dagger}_{\bm{r},\sigma}
- 
\frac{1}{2}
\left\{
\hat{\rho}, \hat{T}^{\dagger}_{\bm{r},\sigma} \hat{T}_{\bm{r}',\sigma'}
\right\}
\right] 
,
\end{equation}
where $\hat{\rho} $ is the density operator.
$\hat{T}_{\bm{r},\sigma} = \sum_{\bm{r}'} \hat{f}^{\dagger}_{\bm{r}',-\sigma} \delta_{\langle \mathbf{R}_{\bm{r}',-\sigma}, \mathbf{R}_{\bm{r},\sigma} \rangle} \hat{f}_{\bm{r},\sigma}$ is the jump operator that leads to displacement of charges toward nearest-neighbor sites, indicated by $\delta_{\langle \mathbf{R}_{\bm{r}',-\sigma}, \mathbf{R}_{\bm{r},\sigma} \rangle}$.
$\mathcal{J}(\bm{R}_{\bm{r},\sigma}-\bm{R}_{\bm{r}',\sigma'})$ and $\Upsilon(\bm{R}_{\bm{r},\sigma}-\bm{R}_{\bm{r}',\sigma'})$ capture respectively the coherent and incoherent couplings.

We start the derivation for the underlying light-matter Hamiltonian of the device presented in the main text, which we again reproduce below as:
\begin{equation}
\label{eq:H}
\hat{H} 
= 
\hat{H}_{\mathrm{hBN}} 
+ 
\hat{H}_{\mathrm{TMD}} 
+ 
\hat{V}
,\quad
\hat{H}_{\mathrm{hBN}} 
= 
\sum_{n} 
\left( 
\nu_{n} - i \frac{\gamma_{n}}{2} 
\right) 
\hat{\alpha}^{\dagger}_{n} \hat{\alpha}_{n}
,\quad
\hat{H}_{\mathrm{TMD}} 
= 
\hat{H}_{\mathrm{M}} 
+ 
\left[ 
\hat{\mathcal{E}}(t) + \hat{\Omega}(t) + \text{H.c.} 
\right]
,
\end{equation}
where $\gamma_n$ denotes the bare decay rate of the HPPs and $\hat{V}$ is expressed in Eq.~\eqref{eq:V}.
The sectors in $\hat{H}_{\mathrm{TMD}}$ are:
\begin{equation}
\label{eq:H_M}
\hat{H}_{\mathrm{M}} 
= 
\sum_{\bm{r},\sigma} 
\left[
(\omega - \omega_f) 
\hat{c}_{\bm{r},\sigma}^{\dagger} 
\hat{c}_{\bm{r},\sigma} 
+ 
\omega_f 
\hat{f}_{\bm{r},\sigma}^{\dagger} 
\hat{f}_{\bm{r},\sigma} 
\right] 
+ 
\hat{V}_{\mathrm{ee}}
,\quad
\hat{V}_{\mathrm{ee}}
= 
- V_B 
\sum_{\bm{r},\sigma}  
\hat{c}_{\bm{r},\sigma}^{\dagger} 
\hat{f}_{\bm{r},\sigma}^{\dagger} 
\hat{f}_{\bm{r},\sigma}  
\hat{c}_{\bm{r},\sigma}  
,
\end{equation}
\begin{equation}
\label{eq:drives}
\hat{\mathcal{E}}(t) 
\simeq 
2 
\mathcal{E} 
e^{-i(\omega - \Delta)t} 
\sum_{\langle\bm{R}_{\bm{r},\sigma},\bm{R}_{\bm{r}', -\sigma}\rangle}
\hat{c}^{\dagger}_{\bm{r},\sigma} \hat{f}^{\dagger}_{\bm{r}', -\sigma}
,\quad
\hat{\Omega}(t) 
= 
\Omega 
e^{-i(\omega - V_B - \Delta - \nu)t} 
\sum_{\bm{r},\sigma} 
\hat{c}^{\dagger}_{\bm{r},\sigma} \hat{f}^{\dagger}_{\bm{r},\sigma}
.
\end{equation}
$\omega_f$ and $\omega$ represent respectively the energy of a valence hole and the bandgap.
$V_B$ is the on-site electron-hole binding strength, wiht its off-site counterpart suppressed for simplicity.
$\mathcal{E}$ and $\Omega$ label respectively the interlayer and intralayer drives, with $\Delta$ and $\Delta + V_B + \nu$ denoting the corresponding detunings.
Note that the above expressions implicitly assume frequency resolution between the drives, dictating $|\Delta|\ll |V_B + \nu|$.

Derivation of Eq.~\eqref{eq:Lindblad_eqn_reprod} from $\hat{H}$ contains four steps, which we outline these steps in respective sections:
First, a polaron transformation is implemented to $\hat{H}$ to derive an effective Hamiltonian (section~\ref{sec:polaron_transform}) in a form that further allows for a rotating-wave approximation (section~\ref{sec:RWA}).
Subsequently, an adiabatic elimination of the conduction-band degrees of freedom is performed (section~\ref{sec:Adiabatic_elimination}), which produces another effective Hamiltonian that is used to obtain the master equation Eq.~\eqref{eq:Lindblad_eqn_reprod} by integrating out the bath (HPP) modes (section~\ref{sec:Born_Markov}).
Finally, in the derivation we neglect several contributions, which we justify in section~\ref{sec:neglect}.

\subsection{Polaron transformation}
\label{sec:polaron_transform}

The polaron (Lang-Firsov~\cite{lang1963kinetic}) transformation used here is subject to the following generator:
\begin{equation}
\hat{S}
=
-
\sum_{\bm{r},\sigma}
\hat{\Pi}_{\bm{r},\sigma}
\hat{f}_{\bm{r},\sigma}^\dagger 
\hat{f}_{\bm{r},\sigma}
,\quad
\hat{\Pi}_{\bm{r},\sigma}
\equiv
\sum_{n}
\frac{g}{\sqrt{2\epsilon_0\nu_n}}
[
\Phi_{n,\sigma}(\bm{r})
\hat{\alpha}_n
-
\Phi_{n,\sigma}^\ast(\bm{r})
\hat{\alpha}_n^\dagger
]
,
\end{equation}
which transforms the Hamiltonian $\hat{H}\to e^{\hat{S}}\hat{H}e^{-\hat{S}}$.
Taking $\gamma_n\ll \nu_n$ and $\frac{g\Phi_{n,\sigma}(\bm{r})}{\sqrt{2\epsilon_0\nu_n}}\ll1$, each sector in the Hamiltonian becomes: 
\begin{equation}
\label{eq:H_undriven_LF_trans}
\hat{H}_{\mathrm{hBN}}
\to 
\hat{H}_{\mathrm{hBN}} 
-
\hat{V}
+
\mathcal{O}\left(\frac{|g\Phi_{\sigma}(\bm{r})|^2}{2\epsilon_0}\right)
,\quad
\hat{H}_{\mathrm{M}}
\to 
\hat{H}_{\mathrm{M}} 
,\quad
\hat{V}
\to
\hat{V}
+
\mathcal{O}\left(\frac{|g\Phi_{\sigma}(\bm{r})|^2}{2\epsilon_0}\right)
,
\end{equation}
\begin{equation}
\label{eq:interlayer_drive_LF_trans}
\hat{\mathcal{E}}(t) 
\to
2 
\mathcal{E} 
e^{-i(\omega - \Delta)t} 
\sum_{\langle\bm{R}_{\bm{r},\sigma},\bm{R}_{\bm{r}', -\sigma}\rangle}
(1-\hat{\Pi}_{\bm{r}', -\sigma})
\hat{c}^{\dagger}_{\bm{r},\sigma} \hat{f}^{\dagger}_{\bm{r}', -\sigma}
+
\mathcal{O}\left(\frac{|g\Phi_{\sigma}(\bm{r})|^2}{2\epsilon_0} \frac{|\mathcal{E}|}{\nu_n}\right)
,
\end{equation}
\begin{equation}
\label{eq:intralayer_drive_LF_trans}
\hat{\Omega}(t) 
\to
\Omega 
e^{-i(\omega - V_B - \Delta - \nu)t} 
\sum_{\bm{r},\sigma} 
(1-\hat{\Pi}_{\bm{r}, \sigma})
\hat{c}^{\dagger}_{\bm{r},\sigma} \hat{f}^{\dagger}_{\bm{r},\sigma}
+
\mathcal{O}\left(\frac{|g\Phi_{\sigma}(\bm{r})|^2}{2\epsilon_0} \frac{|\Omega|}{\nu_n}\right)
,
\end{equation}
where $\mathcal{O}(x)$ indicating corrections that are linear in $x$.
As we will justify in section~\ref{sec:neglect}, it is valid to drop those corrections in the regime of interest.
Combining all contributions up to linear order, we find:
\begin{equation}
\label{eq:H_LF_transform}
\hat{H}
\to
\hat{H}_{\mathrm{hBN}}
+
\hat{H}_{\mathrm{M}}
+
(
\hat{\xi}
+
\mathrm{H.c.}
)
,
\end{equation}
with the effective coupling being:
\begin{equation}
\hat{\xi}
=
e^{-i(\omega - \Delta)t} 
\sum_{\bm{r},\bm{r}'}
\sum_{\sigma,\sigma'}
\xi_{\sigma,\sigma'}^{\bm{r},\bm{r}'}(V_B + \nu)
(1-\hat{\Pi}_{\bm{r}', \sigma'})
\hat{c}^{\dagger}_{\bm{r},\sigma} \hat{f}^{\dagger}_{\bm{r}', \sigma'}
,\quad
\xi_{\sigma,\sigma'}^{\bm{r},\bm{r}'}(x)
\equiv 
2 
\mathcal{E} 
\delta_{\langle\bm{R}_{\bm{r},\sigma},\bm{R}_{\bm{r}', \sigma'}\rangle}
+
e^{i x t} 
\Omega 
\delta_{\bm{r},\bm{r}'}
\delta_{\sigma,\sigma'}
.
\end{equation}
Importantly, all contributions in this coupling are interband, which allows us to implement the standard rotating wave approximation discussed below.

\subsection{Rotating-wave approximation}
\label{sec:RWA}

To make the rotating-wave approximation, we first recast the effective Hamiltonian Eq.~\eqref{eq:H_LF_transform} into:
\begin{equation}
\label{eq:H_for_RWA}
\hat{H}_{\mathrm{hBN}}
+
\hat{H}_{\mathrm{M}}
+
(
\hat{\xi}
+
\mathrm{H.c.}
)
=
\left[ 
\hat{H}_{\mathrm{hBN}} 
+
\hat{H}_{\mathrm{M}} 
- 
\Delta \sum_{\bm{r},\sigma} 
\hat{c}_{\bm{r},\sigma}^\dagger \hat{c}_{\bm{r},\sigma}
\right]
+
\left[
\Delta \sum_{\bm{r},\sigma} 
\hat{c}_{\bm{r},\sigma}^\dagger \hat{c}_{\bm{r},\sigma}
+
(
\hat{\xi}
+
\mathrm{H.c.}
)
\right]
,
\end{equation}
and go to the interaction picture with respect to first term on the right hand side.
This yields:
\begin{equation}
\hat{\xi}
\to
\sum_{\bm{r},\bm{r}'}
\sum_{\sigma,\sigma'}
\xi_{\sigma,\sigma'}^{\bm{r},\bm{r}'}(\nu)
\left(
1
-
\sum_{n}
\frac{g}{\sqrt{2\epsilon_0\nu_n}}
[
\Phi_{n,\sigma'}(\bm{r}')
\hat{\alpha}_n
e^{-i(\nu_n-\frac{i\gamma_n}{2})t}
-
\Phi_{n,\sigma'}^\ast(\bm{r}')
\hat{\alpha}_n^\dagger
e^{i(\nu_n-\frac{i\gamma_n}{2})t}
]
\right)
\hat{c}^{\dagger}_{\bm{r},\sigma} \hat{f}^{\dagger}_{\bm{r}', \sigma'}
.
\end{equation}
Note that the phase factor $e^{iV_Bt}$ in $\xi_{\sigma,\sigma'}^{\bm{r},\bm{r}'}(V_B+\nu)$ is eliminated by the evolution with respect to the interaction in $\hat{H}_{\mathrm{M}}$.
Further assume that $\nu$ is close to some HPP frequencies and drop all fast oscillating terms --- including those $\sim e^{i\nu t} \hat{c}^\dagger
\hat{f}^\dagger$, $\sim e^{i(\nu+\nu_n) t} \hat{c}^\dagger
\hat{f}^\dagger \hat{\alpha}_n^\dagger$, $\sim e^{-i \nu_n t} \hat{c}^\dagger
\hat{f}^\dagger\hat{\alpha}_n$, and $\sim e^{i \nu_n t} \hat{c}^\dagger
\hat{f}^\dagger\hat{\alpha}_n^\dagger$ --- we find:
\begin{equation}
\hat{\xi}
\to
\sum_{\bm{r},\bm{r}'}
\sum_{\sigma,\sigma'}
\left(
2 
\mathcal{E} 
\delta_{\langle\bm{R}_{\bm{r},\sigma},\bm{R}_{\bm{r}', \sigma'}\rangle}
-
\Omega 
\delta_{\bm{r},\bm{r}'}
\delta_{\sigma,\sigma'}
\sum_{n}
\frac{g}{\sqrt{2\epsilon_0\nu_n}}
\Phi_{n,\sigma'}(\bm{r}')
\hat{\alpha}_n
e^{i (\nu-\nu_n + \frac{i\gamma_n}{2}) t}
\right)
\hat{c}^{\dagger}_{\bm{r},\sigma} \hat{f}^{\dagger}_{\bm{r}', \sigma'}
.
\end{equation}
Notably, if $\nu$ is set close to the HPPs with flat dispersion, the mode summation above becomes $\sim\sum_n \Phi_{n,\sigma}(\bm{r}) \hat{\alpha}_n$, which has the interpretation of a local HPP at $\bm{R}_{\sigma}(\bm{r})$.  
This indicates that annhilation of a valence electron (and creation of a conduction electron) is always accompanied with annhilation of a HPP at the same supersite within this polaron rotating frame.

\subsection{Adiabatic elimination of conduction electrons}
\label{sec:Adiabatic_elimination}

Next, we perform adiabatic elimination of the conduction electrons for the effective Hamiltonian in the rotating frame, which is the last term in Eq.~\eqref{eq:H_for_RWA}.
Focusing in the off-resonant regime $|\mathcal{E}|, \frac{|g \Omega \Phi_{\sigma}(\bm{r})|}{\sqrt{2\epsilon_0\nu_n}}, \gamma_n \ll |\Delta|$, this can be achieved by a second order perturbation, which generates three terms that respectively scale with $\sim \frac{|\mathcal{E}|^2}{\Delta}$, $\sim \frac{|\mathcal{E} g \Omega \Phi_{\sigma}(\bm{r})|}{\sqrt{2\epsilon_0\nu_n}\Delta}$, and $\sim \frac{|g \Omega \Phi_{\sigma}(\bm{r})|^2}{2\epsilon_0\nu_n\Delta}$.
We will focus on the middle contribution and drop the other two, which we justify in section~\ref{sec:neglect}.
This yields the following Hamiltonian:
\begin{equation}
\label{eq:H_eff_eliminate_c}
\hat{H}_{\mathrm{eff}}
\simeq
-
\frac{
2 
\mathcal{E}^\ast 
\Omega 
}{\Delta}
\sum_{n,\bm{r}, \sigma}
\frac{g}{\sqrt{2\epsilon_0\nu_n}}
\Phi_{n,\sigma}(\bm{r})
e^{i (\nu-\nu_n + \frac{i\gamma_n}{2}) t}
\hat{T}_{\bm{r}, \sigma}^\dagger
\hat{\alpha}_n
+
\mathrm{H.c.}
,
\end{equation}
recalling that $\hat{T}_{\bm{r},\sigma} = \sum_{\bm{r}'} \hat{f}^{\dagger}_{\bm{r}',-\sigma} \delta_{\langle \mathbf{R}_{\bm{r}',-\sigma}, \mathbf{R}_{\bm{r},\sigma} \rangle} \hat{f}_{\bm{r},\sigma}$, which captures the fermion motion that couples to the HPPs.


\subsection{Integrating out the bath modes}
\label{sec:Born_Markov}

With Eq.~\eqref{eq:H_eff_eliminate_c}, we can now integrate out HPPs within the Born-Markov approximation, which is anticipated to be valid when the system-bath coupling is weak compared to the scale of the bath.
Given that we already assume $\gamma_n\ll \nu_n$ in section~\ref{sec:polaron_transform}, this requirement becomes:
\begin{equation}
\frac{
2 
\mathcal{E}^\ast 
\Omega 
}{\Delta}
\frac{g\Phi_{n,\sigma}(\bm{r})}{\sqrt{2\epsilon_0\nu_n}}
\ll
\gamma_n
,
\end{equation}
which can be achieved by accessing weaker drives or larger detunings.
The resulting Lindbladian master equation is Eq.~\eqref{eq:Lindblad_eqn_reprod}, with the couplings being:
\begin{equation}
\label{eq:J_Upsilon_coeffs}
\begin{bmatrix}
\mathcal{J}(\bm{R}_{\bm{r},\sigma} - \bm{R}_{0,\sigma'})
\\
\Upsilon(\bm{R}_{\bm{r},\sigma} - \bm{R}_{0,\sigma'})
\end{bmatrix}
=
\sum_n
\frac{\nu_n}{2\epsilon_0}
\Phi_{n,\sigma}(\bm{r})
\begin{bmatrix}
\mathrm{Re}(\mathcal{G}_n)
\\
-2
\mathrm{Im}(\mathcal{G}_n)
\end{bmatrix}
\Phi_{n,\sigma'}^\ast(0)
,\quad
\mathcal{G}_n 
\equiv 
\frac{
|
2 
\mathcal{E}^\ast 
\Omega
|^2
}{\Delta^2}
\frac{g^2}{\nu_n^2}
\frac{1}{\nu - \nu_n + \frac{i\gamma_n}{2}}
,
\end{equation}
which reproduces the master equation described in the main text.


\subsection{Justification of dropped terms}
\label{sec:neglect}

Finally, we provide justification to neglect the terms in section~\ref{sec:polaron_transform} and~\ref{sec:Adiabatic_elimination}, which we summarize below:

\begin{enumerate}

\item\label{item_neglect_undriven} Eq.~\eqref{eq:H_undriven_LF_trans}: High order term from polaron transformation to the undriven sector of the Hamiltonian $\sim \frac{|g\Phi_{\sigma}(\bm{r})|^2}{2\epsilon_0}$. The leading correction captures the density-density interaction between the fermions $\sim \hat{f}_{\bm{r},\sigma}^\dagger \hat{f}_{\bm{r},\sigma} \hat{f}_{\bm{r}',\sigma'}^\dagger \hat{f}_{\bm{r}',\sigma'}$ mediated by HPPs in the absence of external drive.

\item\label{item_neglect_interlayer} Eq.~\eqref{eq:interlayer_drive_LF_trans}: Similar to above but from the interlayer drive $\sim \frac{|g\Phi_{\sigma}(\bm{r})|^2}{2\epsilon_0} \frac{|\mathcal{E}|}{\nu_n}$.

\item\label{item_neglect_intralayer} Eq.~\eqref{eq:intralayer_drive_LF_trans}: Similar to above but from the intralayer drive $\sim \frac{|g\Phi_{\sigma}(\bm{r})|^2}{2\epsilon_0} \frac{|\Omega|}{\nu_n}$.

\item\label{item_neglect_adiabatic_E2} Additional term from adiabatic elimination $\sim \frac{|\mathcal{E}|^2}{\Delta}$. The corresponding process describes a fermion that undergoes interlayer transition twice, thereby contribute as $\sim \hat{f}_{\bm{r}',\sigma'}^\dagger \hat{f}_{\bm{r},\sigma}$.

\item\label{item_neglect_adiabatic_Omega2} Another extra term from adiabatic elimination $\sim \frac{|g \Omega \Phi_{\sigma}(\bm{r})|^2}{2\epsilon_0\nu_n\Delta}$. The corresponding process captures a fermion that is locally removed and restored, each coupled to an HPP. Thus, this process contribute as $\sim \hat{f}_{\bm{r},\sigma}^\dagger \hat{f}_{\bm{r},\sigma} \hat{\alpha}_n^\dagger \hat{\alpha}_{n'}$.

\end{enumerate}

Some of them are suppressed because they can be made small compared to the target term Eq.~\eqref{eq:H_eff_eliminate_c}, which scale as $\sim \frac{2 \mathcal{E}^\ast \Omega }{\Delta}\frac{g\Phi_{n,\sigma}(\bm{r})}{\sqrt{2\epsilon_0\nu_n}}$, under certain conditions.
In contrast, the scale of others may not be simultaneously small within the same regime, but still, can be dropped as they are sufficiently short-ranged.
We discuss these considerations respectively in the following sections.

\subsubsection{Suppression due to small scales}

To see how the aforementioned terms can be suppressed by comparing the scales, we first review the assumptions on them that have been made so far:

\begin{itemize}

\item $|\Delta|\ll |V_B + \nu|$: This resolution of frequency is required to distinguish the two drives in Eq.~\eqref{eq:drives}.

\item $\gamma_n\ll\nu_n$: We assume this in section~\ref{sec:polaron_transform} to obtain an anti-Hermitian generator for polaron transformation $\hat{S}$. At the level of implementation, this is also required to frequency-resolve the HPP modes.

\item $\frac{g\Phi_{n,\sigma}(\bm{r})}{\sqrt{2\epsilon_0\nu_n}}\ll1$: This is required in expanding $e^{\hat{S}}$ in series.

\item $|\nu-\nu_n|\ll \nu, \nu_n$: This is required for the rotating-wave approximation discussed in section~\ref{sec:RWA}.

\item $|\mathcal{E}|, \frac{|g \Omega \Phi_{\sigma}(\bm{r})|}{\sqrt{2\epsilon_0\nu_n}}, \gamma_n \ll |\Delta|$: These off-resonant condition is required for adiabatically eliminate the conduction degrees of freedom, discussed in section~\ref{sec:Adiabatic_elimination}.

\item $\frac{2\mathcal{E}^\ast \Omega }{\Delta}\frac{g\Phi_{n,\sigma}(\bm{r})}{\sqrt{2\epsilon_0\nu_n}}\ll\gamma_n$ to validate the Born-Markov approximation discussed in section~\ref{sec:Born_Markov}.

\end{itemize}

There are multiple ways to satisfy the above requirements. 
For instance, the following hierarchy serve as a sufficient condition:
\begin{equation}
\label{eq:hierarchy}
\frac{|\mathcal{E}|}{|\Omega|}
\ll
\frac{g\Phi_{n,\sigma}(\bm{r})}{\sqrt{2\epsilon_0\nu_n}}
\ll
\frac{|\mathcal{E}|}{|\Delta|}
,\,
\frac{|\Omega|}{|\Delta|}
,\,
\frac{\gamma_n}{|\Delta|}
\ll
1
\ll
\frac{\nu_n}{|\Delta|}
\simeq 
\frac{\nu}{|\Delta|}
.
\end{equation}
Note that the first inequality is incorporated to eliminate term \ref{item_neglect_adiabatic_E2} listed in section~\ref{sec:neglect}.
Meanwhile the second inequality removes terms \ref{item_neglect_interlayer} and \ref{item_neglect_intralayer}.
Other contributions, in contrast, cannot be eliminated simply by this hierarchy, and we justify why they are still negligible in the following section.

\subsubsection{Suppression due to locality}

As we will demonstrate in the main text, the combination of HPP modes with $\nu\simeq \nu_n$ that couples to fermions are essentially local.
Such a locality eventually eliminates terms~\ref{item_neglect_undriven} and~\ref{item_neglect_adiabatic_Omega2} listed in section~\ref{sec:neglect}.
Specifically, the correction~\ref{item_neglect_undriven}, is a density-density interaction $\sim \hat{f}_{\bm{r},\sigma}^\dagger \hat{f}_{\bm{r},\sigma} \hat{f}_{\bm{r}',\sigma'}^\dagger \hat{f}_{\bm{r}',\sigma'}$ with the scaling with respect to distance dictated by HPP propagator.
Nevertheless, because HPPs are local, all off-site sectors of such an interaction vanish, and therefore, it contributes only as $\sim (\hat{f}_{\bm{r},\sigma}^\dagger \hat{f}_{\bm{r},\sigma})^2 = \hat{f}_{\bm{r},\sigma}^\dagger \hat{f}_{\bm{r},\sigma}$, which is simply a correction to $\omega_f$.
Similar is true for the other correction~\ref{item_neglect_adiabatic_Omega2}, which captures the process $\sim \hat{f}_{\bm{r},\sigma}^\dagger \hat{f}_{\bm{r},\sigma} \hat{\alpha}_n^\dagger \hat{\alpha}_{n'}$.
The zeroth order contribution from such a process, where $\hat{\alpha}_n^\dagger \hat{\alpha}_{n'}$ is replaced by its expectation value, is also a correction to $\omega_f$.
The next order correction from it provides a density-density interaction between fermions, which vanish again due to the aforementioned short-rangeness.

\section*{Supplementary Note 3 \textbar{} Dissipation matrix from hyperbolic phonon polaritons}

In this section, we elaborate on the spatial dependence of couplings in the Lindbladian expressed in Eq.~\eqref{eq:J_Upsilon_coeffs}. 
For this purpose, explicit expression of $\Phi_{n,\sigma}(\bm{r})$, defined in Eq.~\eqref{eq:phi_W}, is needed.
This further requires the forms of HPP mode functions $\phi_n(\bm{x})$ and electronic orbitals $W_f(\bm{x}-\bm{R}_{\bm{r},\sigma})$, which we discuss below.

The Fröhlich interaction Eq.~\eqref{eq:Frohlich} microscopically originates from the coupling between electric field and the induced dipoles.
Specifically, therein the mode function $\phi_n(\bm{x})\equiv\phi_n(\bm{x},z=\frac{d}{2})$, with $z = \frac{d}{2}$ being the out-of-plane coordinate of the moiré TMD bilayer, can be expressed in terms of the electric field $\bm{E}_n(\bm{x},z)$ as:
\begin{equation}
\bm{E}_n(\bm{x},z)
=
\sqrt{\frac{\nu_n}{2\epsilon_0}}
\bm{F}_n(\bm{x},z)
,\quad
\bm{F}_n(\bm{x},z)
\equiv
-
\nabla
\phi_n(\bm{x},z)
.
\end{equation}
Here $\bm{E}_n(\bm{x},z)$ can be obtained as eigenfunctions of Maxwell equations with the boundary conditions of a slab geometry constituted by hBN within $0\leq z\leq d$, where the HPPs live.
Wwe focus on the according TM modes, which, following Ref.~\cite{feng2026color}, are:
\begin{equation}
\label{eq:field_profile}
\bm{F}_n(\bm{x},z)
=
F_n
e^{i\bm{q}_{||}\cdot\bm{x}}
\begin{cases}
(
\bm{e}_{||}
+
i\bm{e}_{\perp}
)
\mathcal{T}_2 e^{-q_{||}(z-d)}
,\quad
&
z > d
\\
\bm{e}_{||}
(
e^{iq_\perp z}
+
\mathcal{R}
e^{-iq_\perp z}
)
+
\frac{\bm{e}_{\perp}}{\kappa(\nu)}
(
e^{iq_\perp z}
-
\mathcal{R}
e^{-iq_\perp z}
)
,\quad
&
0 \leq z \leq d
\\
(
\bm{e}_{||}
-i
\bm{e}_{\perp}
)
\mathcal{T}_1 e^{q_{||} z}
,\quad
&
z < 0
\end{cases}
,
\end{equation}
recalling that $\bm{q}_{||}$ is the in-plane momenta included in $n$; meanwhile, $q_{\perp} = \frac{q_{\parallel}}{\kappa}$ is the out-of-plane momenta with:
\begin{equation}
\kappa(\nu) 
\simeq 
\sqrt{-\frac{\epsilon^\perp(\nu)}{\epsilon^\parallel(\nu)}}
,\quad
\epsilon_{\parallel}(\nu) 
\simeq
\epsilon_{\parallel}(\infty)
\frac{\nu_{\parallel,\mathrm{LO}}^{2} - \nu^{2}}{\nu_{\parallel,\mathrm{TO}}^{2} - \nu^{2}}
,\quad
\epsilon_{\perp}(\nu) 
\simeq
\epsilon_{\perp}(\infty)
\frac{\nu_{\perp,\mathrm{LO}}^{2} - \nu^{2}}{\nu_{\perp,\mathrm{TO}}^{2} - \nu^{2}}
,
\end{equation}
with $\nu_{\parallel,\mathrm{LO}}$ and $\nu_{\parallel,\mathrm{TO}}$ denoting the frequencies of longitudinal and transversal op-
tical phonons along the in-plane direction, and $\nu_{\perp,\mathrm{LO}}$ and $\nu_{\perp,\mathrm{TO}}$ are the out-of-plane counterpart.
The boundaries are matched by the following transmission and reflectinon coefficients:
\begin{equation}
\mathcal{R}
=
\frac{i\epsilon_{||}(\nu)\kappa(\nu) + 1}{i\epsilon_{||}(\nu)\kappa(\nu) - 1}
,\quad
\mathcal{T}_1
=
1
+
\mathcal{R}
,\quad
\mathcal{T}_2
=
e^{iq_\perp d}
+
\mathcal{R}
e^{-iq_\perp d}
,\quad
\mathcal{R}^2
=
e^{2iq_\perp d}
,
\end{equation}
where we set the dielectric constant of exterior regions to be $1$.
The overall amplitude $F_n$ is set by the normalization condition~\cite{archambault2010quantum,rivera2020light,feng2026color}:
\begin{equation}
\frac{1}{2\nu_n}
\int d^2\bm{x} dz
\,
\bm{F}_n^\ast(\bm{x},z)
\cdot
\frac{d(\nu^2\bm{\epsilon}(\nu))}{d\nu}\bigg|_{\nu = \nu_n}
\cdot
\bm{F}_n(\bm{x},z)
=
1
,\quad
\bm{\epsilon}(\nu)
= 
\epsilon_{||}(\nu)
\bm{e}_{||}
\otimes
\bm{e}_{||}
+
\epsilon_{\perp}(\nu)
\bm{e}_{\perp}
\otimes
\bm{e}_{\perp}
.
\end{equation}
Plugging in the above mode profile, the normalization condition becomes:
\begin{equation}
|F_n|
=
\sqrt{
\frac{2}{I L^2 d}
}
,\quad
I 
= 
I_{\text{hBN}} 
+
I_{\text{ext}} 
+
I'_{\text{hBN}}
+
I'_{\text{ext}} 
.
\end{equation}
, where $I$ is the dimensionless version of the above integral and $L^2$ denotes the area of the plane.
We further separate into four contributions, depending on whether they originate from hBN slab and exterior regions, and on whether they involve the dielectric tensor (unprimed) or its derivative (primed).
Upon this treatment, one can show that $I'_{\text{ext}} = 0$ as the dielectric constant outside of hBN is not frequency dependent, and that $I_{\text{ext}} = -I_{\text{hBN}}$ after plugging in the expressions of the reflection and transmission coefficients.
The remaining contribution is thus:
\begin{equation}
I'_{\text{hBN}} 
=
(2\nu_n)
\left[
\left(
1
+
\frac{\text{Im}[\mathcal{R}]}{q_\perp d}
\right)
\dot{\epsilon}_{||}(\nu_n)
+
\left(
1
-
\frac{\text{Im}[\mathcal{R}]}{q_\perp d}
\right)
\frac{\dot{\epsilon}_{\perp}(\nu_n)}{|\kappa(\nu)|^2}
\right]
.
\end{equation}
With $|F_n|=\sqrt{\frac{2}{I'_{\text{hBN}} L^2 d}}$, we can obtain the HPP mode function and the corresponding electric field at the coordinate of the bilayer $z = \frac{d}{2}$ as:
\begin{equation}
\label{eq:phi_x}
\phi_n(\bm{x}) 
= 
-
\frac{1}{iq_{||}} 
\sqrt{\frac{2\epsilon_0}{\nu_n}}
E_n^{||}\left(\bm{x},\frac{d}{2}\right)
,\quad
E_n^{||}\left(\bm{x},\frac{d}{2}\right)
=
\sqrt{\frac{\nu_n}{2\epsilon_0}}
F_n
e^{i\bm{q}_{||}\cdot \bm{x}}
(
e^{\frac{iq_\perp  d}{2}} + \mathcal{R}\, e^{-\frac{iq_\perp  d}{2}}
)
.
\end{equation}




The amplitude $\Phi_{n,\sigma}(\bm{r})$ can be obtained by weighting the above expression with the electronic distribution $|W_f(\bm{x}-\bm{R}_{\bm{r},\sigma})|^2$ then integrating over $\bm{x}$.
This electronic weight would provide a length scale that confines the relevant momenta range that eventually contributes to Eq.~\eqref{eq:J_Upsilon_coeffs}.
To demonstrate this point, we approximate the electronic orbital with the following Gaussian profile:
\begin{equation}
W_f(\bm{x}-\bm{R}_{\bm{r},\sigma})
=
\frac{1}{\sqrt{\pi} a_W}
e^{-\frac{(\bm{x}-\bm{R}_{\bm{r},\sigma})^2}{2a_W^2}}
,
\end{equation}
whose width is captured by the length scale $a_W$.
Plugging this and Eq.~\eqref{eq:phi_x} into Eq.~\eqref{eq:phi_W}, we find:
\begin{equation}
\label{eq:Phi_Gaussian}
\Phi_{n,\sigma}(\bm{r})
=
\varphi_{\bm{q},l}
\exp\left[
i\bm{q}\cdot\bm{R}_{\bm{r},\sigma}
-
\frac{a_W^2 \bm{q}^2}{4}
\right]
,\quad
\varphi_{\bm{q},l}
=
-
\frac{1}{iq_{||}} 
\sqrt{\frac{2\epsilon_0}{\nu_n}}
E_n^{||}\left(0,\frac{d}{2}\right)
.
\end{equation}

\subsection{Integration over all modes}

Plugging in all these expressions back into Eq.~\eqref{eq:J_Upsilon_coeffs}, we find:
\begin{equation}
\Upsilon(\bm{R})
\sim
\sum_{\bm{q}_{||},l}
\frac{|F_n|^2}{q_{||}^2}
e^{i\bm{q}_{||}\cdot\bm{R}-\frac{a_W^2 q_{||}^2}{2}}
\frac{\gamma_n}{\nu_n}
\frac{
1
+
\text{Re}
[
\mathcal{R}e^{-iq_\perp d}
]
}{(\nu-\nu_n)^2 + \frac{\gamma_n^2}{4}}
,
\end{equation}
where the spatially independent proportionality constant is suppressed here for simplicity.
To further simplify the above expression, we approximate $\bm{q}_{||}$ as a continuous variable, which is implemented by replacing $\sum_{\bm{q}_{||}}$  with $\frac{L^2}{(2\pi)^2}\int d\theta dq_{||} \,q_{||}$ , giving:
\begin{equation}
\Upsilon(\bm{R})
\sim
\sum_{l}
\int dq_{||}
\frac{J_0(q_{||}R)}{q_{||}}
\frac{e^{-\frac{a_W^2 q_{||}^2}{2}}}{I'_{\text{hBN}} }
\frac{\gamma_n}{\nu_n}
\frac{
1
+
\text{Re}
[
\mathcal{R}e^{-iq_\perp d}
]
}{(\nu-\nu_n)^2 + \frac{\gamma_n^2}{4}}
,
\end{equation}
where $J_0(x)$ denotes the Bessel function.
Numerical evaluation of this integral allows for extracting the scaling of $\Upsilon(\bm{R})$ with distance $R$.


\section*{Supplementary Note 4 \textbar{} Collective layer-sublattice transition}

In this section, we elaborate on how to implement the master equation providing superradiant burst with the same moiré heterostructure.
To reiterate, this model reads:
\begin{equation}
\label{eq:master_eq_SR}
\partial_t \hat{\rho} 
= 
-i 
J 
[\hat{T}^\dagger \hat{T}, \hat{\rho}] 
+ 
\Gamma 
\left( \hat{T} \hat{\rho} \hat{T}^\dagger - \frac{1}{2} \{ \hat{T}^\dagger \hat{T}, \hat{\rho} \} \right)
,
\end{equation}
with $J$ and $\Gamma$ captures the collective coherent and dissipative coupling given by light-matter interaction wihtin Born-Markov approximation.
The corresponding jump operator reads:
\begin{equation}
\label{eq:jump_SR}
\hat{T}
=
\sum_{\bm{k}}
w_-
\eta_{\bm{k}}^\ast
\hat{\mathcal{S}}_{\bm{k}}^{-}
+
w_+
\eta_{\bm{k}}
\hat{\mathcal{S}}_{\bm{k}}^{+}
,\quad
\eta_{\bm{k}}
=
\sum_{\bm{\delta}}
e^{i\bm{k}\cdot\bm{\delta}}
,\quad
\hat{\mathcal{S}}_{\bm{k}}^{\pm}
=
\hat{f}_{\bm{k},\pm}^\dagger
\hat{f}_{\bm{k},\mp}
,
\end{equation}
Here $\{\bm{\delta}\}$ are the set of vectors pointing from a $\sigma = -$ site to its nearest neighbors.
$\hat{f}_{\bm{k},\sigma}$ is the momentum $(\bm{k})$ space representation of $\hat{f}_{\bm{r},\sigma}$.
$w_\sigma$ is the layer-sublattice weight on the dissipative matrix, as we elaborate later.

To begin with, we introduce a layer bias by applying an out-of-plane displacement field.
Accordingly, in the matter Hamiltonian Eq.~\eqref{eq:H_M}, the cost of generating a hole (in both conduction and valence bands) is shifted oppositely for the two layers, i.e. $\omega_f\to \omega_f + \sigma \omega_b$, with $\omega_b$ denoting the shift for the $\sigma = +$ layer. 
Due to this bias, the interlayer transition of a valence electron at layer $\sigma$ to conduction band at layer $-\sigma$ costs an energy $\omega \to \omega + 2\sigma \omega_b$.
Such a difference between the two interlayer transitions allow us to access them with different driving frequencies.
Here we consider the following forms of these drives that are slightly modified from Eq.~\eqref{eq:drives}:
\begin{equation}
\hat{\mathcal{E}}(t) 
\to
\sum_{\langle\bm{R}_{\bm{r},+},\bm{R}_{\bm{r}', -}\rangle} 
\mathcal{E}_+
e^{-i(\omega - V_B - 2 \omega_b- \Delta)t} 
\hat{c}^{\dagger}_{\bm{r},+} \hat{f}^{\dagger}_{\bm{r}', -}
+
\sum_{\langle\bm{R}_{\bm{r},-},\bm{R}_{\bm{r}',+}\rangle}
\mathcal{E}_-
e^{-i(\omega - V_B + 2\omega_b - \Delta)t} 
\hat{c}^{\dagger}_{\bm{r},-} \hat{f}^{\dagger}_{\bm{r}',+}
,
\end{equation}
where the frequencies are chosen such that they respectively cancel out the layer bias later in the rotating frame.

Meanwhile, here we completely suppress the intralayer drives.
This is because we want to access emission of photons (instead of HPPs) at frequencies of order the bandgap, since the corresponding wavelength ($\simeq800$ nm) is much large than the moiré lattice spacings (typically $10$ nm), thereby naturally offer non-local dissipation~\cite{huang2025collective}.
To capture the essential physics with simplicity, hereafter we assume the ratio between these length scales approaches zero, namely, the moiré lattice is deep-subwavelength.
As a consequence, only light with zero momentum couples to the transition.

Based on the above considerations, we make the following change to several sectors of the model Hamiltonian captured in Eq.~\eqref{eq:H}:
\begin{equation}
\hat{H}_{\mathrm{hBN}} 
\to
\hat{H}_{\mathrm{L}}
=
\omega_L
\hat{b}^\dagger
\hat{b}
,\quad
\hat{V}
\to
\hat{V}_{LM}
\simeq
g_{LM}
\hat{b}
\hat{X}^\dagger
+
g_{LM}^\ast
\hat{b}^\dagger
\hat{X}
,\quad
\hat{X}
=
\sum_{\bm{r},\sigma}
\hat{c}_{\bm{r},\sigma} 
\hat{f}_{\bm{r},\sigma}
,\quad
g_{LM}
=
\sqrt{\frac{\omega_L}{2\epsilon_0\mathcal{V}}}
\mathcal{D}
,
\end{equation}
where $\hat{b}$ denotes the annihilation operator of a zero-momentum photon, and its frequency is $\omega_L$.
$\mathcal{V}$ is the mode volume, and $\mathcal{D}$ labels the optical matrix element for the intralayer transition.
Notably, here we suppress for simplicity the effect of dynamical screening from hBN at the photonic frequency, such that the permittivity is still the value obtained from vacuum, $\epsilon_0$.
Meanwhile, as in the main text, here we assume that one (real) spin species of electronic degrees of freedom is gapped out by external magnetic field, and therefore, only the other is relevant.
Due to the selection rules in TMDs~\cite{huang2023mott}, the corresponding transition only couples to one (circular) polarization of light.
In the above expressions, we thus omit those labels for simplicity.

Next, we integrate out the photons within Born-Markov approximation, which is valid the light-matter coupling $g_{LM}$ is slow compared to $\omega_L$.
Specifically, $\hat{H}_{L} + \hat{V}_{LM}$ provides the following Lindblad operator:
\begin{equation}
\hat{D}_L[\hat{\rho}]
=
\gamma_L
\left(
\hat{X}
\hat{\rho}
\hat{X}^\dagger
-
\frac{1}{2}
\{
\hat{\rho}
,
\hat{X}^\dagger
\hat{X}
\}
\right)
,\quad
\gamma_L
=
\frac{\mathcal{D}^2\omega_L^3}{3\pi\epsilon_0 c^3}
,
\end{equation}
and the associated coherent Hamiltonian $\sim \hat{X}^\dagger \hat{X}$.
Upon adiabatical elimination of the conduction band degrees of freedom with $|\mathcal{E}_{\sigma}|\ll|\Delta|$, they yield the master equation Eq.~\eqref{eq:master_eq_SR} with $\Gamma \sim \frac{|\mathcal{E}_+|^2 + |\mathcal{E}_-|^2}{\Delta^2}\gamma_L$ (and similar rescaling is present in $J$) and the dressed jump operator:
\begin{equation}
\hat{T}
=
\sum_{\langle\bm{R}_{\bm{r},+},\bm{R}_{\bm{r}',-}\rangle}
w_-
\hat{f}_{\bm{r}', -}^\dagger
\hat{f}_{\bm{r}, +}
+
\sum_{\langle\bm{R}_{\bm{r},-},\bm{R}_{\bm{r}',+}\rangle}
w_+
\hat{f}_{\bm{r}', +}^\dagger
\hat{f}_{\bm{r}, -}
,\quad
w_{\sigma}
=
\frac{\mathcal{E}_{-\sigma}}{\sqrt{|\mathcal{E}_+|^2 + |\mathcal{E}_-|^2}}
,
\end{equation}
which becomes Eq.~\eqref{eq:jump_SR} upon Fourier transformation to momentum space.

\subsection{The role of coherent Hamiltonian on superradiant burst}

In the main text, we numerically demonstrate that the dynamics purely from the Lindbladian Eq.~\eqref{eq:master_eq_SR} with $w_+\to 0$ exhibits a superradiant burst at early times.
Therein, the coherent Hamiltonian $J \hat{X}^\dagger \hat{X}$ is not included as it typically provides fast oscillations leading to numerical instability.
However, in the rest of this section, we analytically demonstrate that including this coherent term would not affect the presence of the burst, despite that it could potentially affect the dynamics qualitatively.

The key of the following proof is to use the fact that the emission rate $\frac{d}{dt}\sum_{\bm{k}} \langle \hat{f}_{\bm{k}, +}^\dagger \hat{f}_{\bm{k}, +}\rangle$ approaches zero at long times.
Therefore, following extreme value theorem, a local maximum in such rate --- signaling the burst --- must exist if initially its derivative is positive.
The statement implies that if $\frac{d^2}{dt^2}\sum_{\bm{k}} \langle \hat{f}_{\bm{k}, +}^\dagger \hat{f}_{\bm{k}, +}\rangle\big|_{t=0}$ is independent of $J$, the existence of the burst is irrelevant of the participation of the coherent Hamiltonian.

The remaining of the proof is to compute the above second derivative at initial time.
To do this, we first evaluate the first derivative as:
\begin{equation}
\frac{d}{dt}
\sum_{\bm{k}} 
\langle 
\hat{f}_{\bm{k}, +}^\dagger 
\hat{f}_{\bm{k}, +}
\rangle
=
\Gamma
\sum_{\bm{k}} 
\langle 
\hat{T}^\dagger
\hat{f}_{\bm{k}, +}^\dagger 
\hat{f}_{\bm{k}, +}
\hat{T}
-
\hat{T}^\dagger
\hat{T}
\hat{f}_{\bm{k}, +}^\dagger 
\hat{f}_{\bm{k}, +}
\rangle
,
\end{equation}
where we use the fact that $\hat{T}^\dagger \hat{T}$ commutes with the total fermion number to eliminate the contribution from the coherent term and to move the operators in the incoherent terms.
Simplifying the right-hand side, we find (for $w_+\to 0$):
\begin{equation}
[
\hat{T}
,
\hat{f}_{\bm{k}, +}^\dagger 
\hat{f}_{\bm{k}, +}
]
=
w_-
\eta_{\bm{k}}^\ast
\hat{f}_{\bm{k}, -}^\dagger 
\hat{f}_{\bm{k}, +}
\Longrightarrow
\frac{d}{dt}
\sum_{\bm{k}} 
\langle 
\hat{f}_{\bm{k}, +}^\dagger 
\hat{f}_{\bm{k}, +}
\rangle
=
-
\Gamma
\langle 
\hat{T}^\dagger
\hat{T}
\rangle
.
\end{equation}
Note that the right-hand side is proportional to the coherent term $J\hat{T}^\dagger \hat{T}$. 
Therefore, in the next derivative, the corresponding contribution $\sim J\Gamma[\hat{T}^\dagger \hat{T}, \hat{T}^\dagger \hat{T}]$ vanishes.
Specifically, we find:
\begin{equation}
\frac{d^2}{dt^2}
\sum_{\bm{k}} 
\langle 
\hat{f}_{\bm{k}, +}^\dagger 
\hat{f}_{\bm{k}, +}
\rangle
=
-
\Gamma
\langle 
\hat{T}^\dagger
\hat{T}^\dagger
\hat{T}
\hat{T}
-
\hat{T}^\dagger
\hat{T}
\hat{T}^\dagger
\hat{T}
\rangle
.
\end{equation}
Notably, while the right-hand side could possess an implicit $J$ dependence encoded in the time-evolved state, such a dependence is absent at $t = 0$, since the initial state is chosen irrespective of $J$.
Therefore, the second derivative at initial time is $J$ independent, and following the above arguments, a superradiant burst must exist.

\bibliography{Biblio}